\documentclass[useAMS,usenatbib]{mn2e}
\usepackage{epsfig}

\def\ML{\mbox{$M/L$}} 
\def\dML{\mbox{$\nabla_{\ell} \Upsilon$}}
\def\dMLo{\mbox{$\nabla_{\ell}^{\rm obs} \Upsilon$}}
\def\dMLm{\mbox{$\nabla_{\ell}^{\rm mod} \Upsilon$}}

\newcommand\gsim{\mathrel{\raise.3ex\hbox{$>$}\mkern-14mu
             \lower0.6ex\hbox{$\sim$}}}
\newcommand\lsim{\mathrel{\raise.3ex\hbox{$<$}\mkern-14mu
             \lower0.6ex\hbox{$\sim$}}}

\title[Mass-to-light ratio gradients in early-type galaxy haloes]{Mass-to-light ratio gradients in early-type galaxy haloes}
\author[Napolitano et al.]{N.R. Napolitano$^{1}$\thanks{E-mail:
nicola@astro.rug.nl}, M. Capaccioli$^{2,3}$,
A.J. Romanowsky$^{4,5}$, N.G. Douglas$^{1}$, 
\and M.R. Merrifield$^{5}$, 
K. Kuijken$^{1,6}$, M. Arnaboldi$^{7}$, O. Gerhard$^{8}$,  K.C. Freeman$^{9}$\\
$^{1}$Kapteyn Astronomical Institute, Postbus 800, 9700 AV Groningen, The Netherlands\\
$^{2}$INAF--Astronomical Observatory of Capodimonte, via Moiariello
16, I-80131 Naples, Italy\\
$^{3}$Dept. of Physical Sciences, University ``Federico II'', Naples,
Italy\\
$^{4}$Departamento de F\'{i}sica, Universidad de Concepci\'{o}n, Casilla 160-C, Concepci\'{o}n, Chile\\
$^{5}$School of Physics and Astronomy, University of Nottingham, University Park, Nottingham NG7 2RD, England\\
$^{6}$Leiden Observatory, Niels Bohrweg 2, NL-2333 CA Leiden, The Netherlands\\
$^{7}$INAF--Astronomical Observatory of Pino Torinese, via
Osservatorio 20, I-10025 Pino Torinese, Italy\\
$^{8}$Astronomy department, University of Basel, Venusstrasse, Basel, Switzerland\\
$^{9}$RSAA, Mt. Stromlo Observatory, Weston Creek P.O., ACT 2611, Australia}

\begin{document}

\date{Accepted . Received ; in original form }

\pagerange{\pageref{firstpage}--\pageref{lastpage}} \pubyear{2004}

\maketitle

\label{firstpage}

\begin{abstract}
Since the near future should see a rapidly expanding set of probes of the halo masses of
individual early-type galaxies,
we introduce a convenient parameter for characterising the halo masses from both
observational and theoretical results:
\dML, the logarithmic radial gradient of the mass-to-light ratio.
Using halo density profiles from $\Lambda$CDM simulations,
we derive predictions for this gradient for various galaxy luminosities and
star formation efficiencies $\epsilon_{\rm SF}$.
As a pilot study, we
assemble the available \dML\ data from kinematics in early-type galaxies---representing
the first unbiassed study of halo masses in a wide range of early-type galaxy luminosities---and
find a correlation between luminosity and \dML, such that the brightest
galaxies appear the most dark-matter dominated.
We find that the gradients in most of the brightest galaxies may fit in well with
the $\Lambda$CDM predictions, 
but that there is also a population of fainter galaxies whose gradients are so low
as to imply an unreasonably high star formation efficiency $\epsilon_{\rm SF} > 1$.
This difficulty is eased if dark haloes are not assumed to have the standard
$\Lambda$CDM profiles, but lower central concentrations.

\end{abstract}

\begin{keywords}
galaxies: haloes, fundamental parameters, evolution, kinematics and dynamics -- dark matter
\end{keywords}

\section{Introduction} 

Spiral galaxies have long been known to show apparent increases in the mass-to-light ratio (\ML) in their outer parts,
which was one of the prime factors in discovering the ubiquity of dark matter in the universe
\citep{BvdK79,WFR88}.
Current studies of spiral galaxy rotation curves largely focus on the observed radial distribution of the dark matter
vis-\`{a}-vis the predictions of the current cosmological paradigm, cold dark matter with a cosmological constant
($\Lambda$CDM).
In particular, $\Lambda$CDM simulations of galaxy halo formation predict a steep cusp of dark matter in the halo
centres (\citealt{nfw96,nfw97}: NFW hereafter; \citealt{moore99}),
while many observations of late-type galaxies indicate relatively low central concentrations of dark matter
(see \citealt{alam,march,dB03,gentile04}; and references therein).
It remains to be seen if this discrepancy can be traced to observational problems, to oversimplified 
predictions of the halo properties (especially with respect to the inclusion of baryonic effects), 
or to a fundamental failure of the $\Lambda$CDM paradigm.

Early-type galaxies (ellipticals and lenticulars) are as numerous as late-types, and it is important to see how their
dark matter distributions compare to those of spirals and to theoretical predictions.
Historically,
the lack of easily interpretable mass tracers like the cold gas in spirals has 
made it very difficult to map the mass distribution in early-types
at large distances from their centres.
Observations of stellar kinematics from integrated-light spectra are mostly confined within
$2R_{\rm e}$ (where the effective radius $R_{\rm e}$ encloses half the projected light), 
which is generally insufficient for establishing the presence of a dark matter halo \citep{K00,MB01}---much less
for determining its detailed properties.
Attempts to constrain the dark matter content in this way have produced inconsistent results
\citep{gerhard01,BSD03,tbb04,pad04,ML04a},
demonstrating the difficulties inherent in studying the stellar-dominated regions.

Tracers at larger radius in early-type galaxies are needed, and
recent advances in instrumentation have dramatically improved the feasibility of
measuring halo masses using 
planetary nebulae (PNe; \citealt{dou02}), globular clusters (GCs; \citealt{bd04}), gaseous X-ray emission \citep{km02,OsPon04b}, gravitational lensing \citep{Ket01}, and extended HI disks \citep{Oost02}.
Such studies have so far ascertained the presence of both massive \citep{rk01,db02,nap02,pc03,tr04,OsPon04b,TrKoo04}
and weak \citep{mend,ral03,peng04} dark haloes.
Various attempts have been made using halo tracers
to derive generic constraints on dark matter in ellipticals
\citep{bertola,bahcall,danz,lw99},
but none of these has used a large, well-defined sample of ellipticals covering
a broad range of luminosities.
With ongoing observational projects aimed to the study of mass distribution around early-type galaxies using stellar kinematics (SAURON: \citealt{SAUR}; PN.S project: \citealt{dou02}),
strong gravitational lensing (CLASS: \citealt{CLAS}; and LSD: \citealt{LSD})
and X-rays \citep{OsPon04a,OsPon04b}, we expect
to have, in the near future, a much better observational picture of the halo mass distributions in a large representative sample of early-type galaxies.

To prepare for this observational onslaught, 
in this paper we want to provide a suite of predictions from the $\Lambda$CDM theory for halo mass distributions.
To gain qualitative insight, we begin with the most basic characterisation of
a dark matter halo around a luminous galaxy: the \ML\ radial gradient.
In conjunction with deriving predictions for this gradient, 
we illustrate the kind of analysis possible with future extensive data sets
by comparing the currently available observational results with the model expectations.
This approach may be contrasted with the detailed analysis of individual galaxies,
where one attempts to decompose the mass distributions
into their baryonic and non-baryonic components
and then compare these to $\Lambda$CDM predictions.
Here we are more simply looking at a broad property of the mass profiles 
which can be statistically analysed.
Note that even with detailed analyses, 
it may never be possible in individual
galaxies to unravel all the geometrical and dynamical degeneracies,
necessitating a statistical approach such as in this paper
in order to interpret the results.

The paper is organised as follows: 
in Section 2 we introduce the \ML\ gradient \dML\
as a basic quantity for describing the relative distributions of luminous and dark mass;
in Section 3 we present predictions from $\Lambda$CDM for \dML,
including its dependence on various galaxy properties;
in Section 4 we assemble observational results on \dML\ for a sample of galaxies from the literature,
and we test for correlations with other observed properties;
in Section 5
we compare the observed and predicted trends in \dML,
and examine the implications for galaxy formation.
We draw conclusions in Section 6.

\section{The \ML\ gradient}
\label{mlgrad}

Aiming for the simplest characterisation of the dark matter content of a galaxy, we could consider
the total \ML\ at some large radius,
such as the benchmark quantity $\Upsilon_{{\it B}5}$ ($B$-band $M/L$ within 5~$R_{\rm e}$) introduced
by \cite{ral03}.
This would allow for constructions similar to the Tully-Fisher relations in spiral galaxies, but
in comparing different galaxies, a quantity such as $\Upsilon_{{\it B}5}$
is complicated by differences in the stellar luminosities (i.e., in the stellar $M/L$, $\Upsilon_*$).
Instead, we consider the increase in the \ML\ with radius relative to its value in the galaxy centre
(where it is presumably dominated by $\Upsilon_*$).
This \ML\ gradient gives us a clearer idea of how much extra mass resides in the halo,
and it is computed below.

The \ML\ gradient is a continuously varying quantity with the (three-dimensional) radius, 
but its average value between an inner and an outer radius ($r_{\mathrm{in}}, r_{\mathrm{out}}$) 
is given by:
\begin{equation} 
\small
\frac{\Delta\Upsilon}{\Delta r}=\frac{\Upsilon_{\mathrm{out}}-\Upsilon_{\mathrm{in}}}{r_{\mathrm{out}}-r_{\mathrm{in}}}=
\frac{\Upsilon_*}{\Delta r}\left[ 
\left(\frac{M_{\rm d}}{M_*} \right)_{\mathrm{out}} - \left(\frac{M_{\rm d}}{M_*} 
\right)_{\mathrm{in}}\right] ,
\label{grad0} 
\end{equation}
where $M_*(r)$ and $M_{\rm d}(r)$ are respectively the luminous (stellar) and the dark masses enclosed within a certain radius. 
This is the absolute value of the gradient, but to normalise galaxies of different mass and size to
similar scales, we compute
\begin{equation} 
\small
\frac{R_{\rm e} {} \Delta\Upsilon}{\Upsilon_*~\Delta r}=\frac{R_{\rm e}}{\Delta r}\left[ 
\left(\frac{M_{\rm d}}{M_*} \right)_{\mathrm{out}} - \left(\frac{M_{\rm d}}{M_*} 
\right)_{\mathrm{in}}\right] 
\equiv \dML .
\label{grad}
\end{equation}

Thus \dML\ is approximately equivalent to the logarithmic gradient of the \ML.
The predicted value for \dML\ is the middle expression of Eq. \ref{grad} (also referred to as \dMLm\ hereafter),
and is uniquely determined once the mass distributions 
and measurement radii are adopted (see Section~\ref{mod}).
The empirical value is the left expression (\dMLo, hereafter); in addition to any dynamical modelling uncertainties in $\Delta \Upsilon$,
$\Upsilon_*$ is a source of systematic uncertainty (discussed later).
Although \dML\ will in principle depend on the location of its measurements ($r_{\mathrm{in}}/R_{\rm e}$, $r_{\mathrm{out}}/R_{\rm e}$),
the ratio $M_{\rm d}(r)/M_*(r)$ can, to a first approximation, be assumed to vary linearly with the radius
(i.e., not to have strong local gradients),
such that \dML\ is fairly independent of the measurement radius
(we check this for a theoretical model in Sections~\ref{frac} and~\ref{pred}; see also \citealt{dub98}).
This allows us to uniformly compare observational results where the values of 
$(r_{\rm in}/R_{\rm e}, r_{\mathrm{out}}/R_{\rm e})$ are
rather heterogeneous.
Note also that \dML\ is independent of the filter bandpass used for the \ML\ estimates.

\section{Predictions from $\Lambda$CDM}
\label{mod}

We build spherical representations of early-type galaxy mass profiles, using a constant-$M/L$
model for the stellar distribution (Section~\ref{lum}) 
plus a $\Lambda$CDM model of the dark halo (Section~\ref{cdm}).
We combine these components in Section~\ref{frac}.
We do not include a diffuse gas component, since
we expect its mass inside $\sim$5$R_{\rm e}$ to be a small fraction of the stellar mass
(we discuss the effect of this assumption in Section~\ref{pred}).
We derive the predictions for various model parameters in Section~\ref{pred}.

\subsection{The luminous component}
\label{lum}
The \cite{hern90} profile is 
known to be a fairly accurate representation of the stellar mass distribution in early-type
galaxies; in projection, it approximates well
the de Vaucouleurs  $R^{1/4}$ law \citep{dV48}. 
Assuming $\Upsilon_*$ to be radially constant, the stellar mass density distribution is
\begin{equation}
\rho_*(r)=\frac{M_*}{2\pi} \frac{k\ R_{\rm e}}{r\ (r+k\ R_{\rm e})^3} ,
\label{rhostar}
\end{equation}
where 
$k \simeq 0.5509$. 
The cumulative mass profile is
\begin{equation}
M_*(r)=M_* \frac{(r/R_{\rm e})^2}{(r/R_{\rm e}+k)^2}.
\label{Mstar}
\end{equation}
Thus we have a family of density distributions characterised by two parameters:
the total luminous mass, $M_*$, and the effective radius $R_{\rm e}$. 
These parameters do not vary arbitrarily in early-type galaxies (as demonstrated by the fundamental plane (FP)) but are correlated by a power law of the form
\begin{equation}
\frac{R_{\mathrm{e}}}{h^{-1} {\rm kpc}}=R_M \times \left(\frac{M_{\mathrm{*}}}{h^{-1}10^{11} M_{\odot}}\right)^{\alpha}.
\label{remleq}
\end{equation}
The scatter of the parameters $(\alpha, R_M)$ in Eq.~\ref{remleq} 
is an important source of uncertainty which we will address in more detail in later sections.
In this section we start drawing predictions for generic early-type galaxies in $\Lambda$CDM. For this reason we assume the results from \citet{shen03} based on a large statistical sample (about 140\,000 galaxies) as a reasonable representation of the size-mass scaling properties of early-types.\\
\citet{shen03} found for their SDSS sample of early-type galaxies that $\alpha=0.56$ and $R_M=3.6$ ($R_M= 4.17$ for $h=0.7$), 
consistent also with the results of \cite{ch98}.
Other studies of large samples of early-type galaxies have found mean relations between total luminosity and effective radius of
$R_{\rm e}\propto L^{\gamma}$, with $\gamma=$~0.54--0.63 \citep{pahre98,bernardi03,ML04a},
i.e. a slope very similar to the \cite{shen03} results under the assumption that $\Upsilon_*$ is constant with $M_*$ (and $L$). In this case we could take $\alpha=\gamma=0.6$ as an average value.
If this assumption is not true, as suggested by FP studies,
then a relation $\Upsilon_* \propto M_*^\beta$ implies $R_{\rm e}\propto L^{\gamma} \propto M_*^{(1-\beta)\gamma}$ 
and $\alpha=(1-\beta)\gamma$. 
The exponent $\beta$ depends on the passband and 
has been found to lie in the range $\beta \sim$~0.2--0.6 \citep{pahre95,gerhard01};
assuming $\gamma=0.6$ and an intermediate value of $\beta=0.3$, we obtain $\alpha=0.4$.
Higher values are found using
the $\mu_{\mathrm{e}}-R_{\mathrm{e}}$ relation \citep{kor77}
and the assumption of homology
($L \propto I_{\mathrm{e}} R_{\mathrm{e}}^2$, 
where $\mu_{\rm e}$ and 
$I_{\mathrm{e}}$ are the effective surface brightness in magnitudes and luminosities, respectively).
The Kormendy relation $\mu_e\sim 3 \log R_{\rm e}$ implies $\log I_{\rm e}\sim -1.2 \log R_{\rm e}$ and with a bit of algebra, $\gamma=1.25$,
giving $\alpha=$~0.9--1.2 for $\beta=$~0--0.3.\\
We thus adopt $\alpha=0.56$ as a characteristic value in equation 5, with
$\alpha=0.4$--$1.2$ representing the plausible range.
Later in the paper,
when comparing model predictions with observations, we will check that
this size-mass relation is consistent with our galaxy sample (Section
4.2). Note that, besides the relation $(\alpha=0.56, ~R_M=3.6)$ for massive galaxies ($M_*>2\times 10^{10} M_{\odot}$), \citet{shen03} also provide a shallower relation ($\alpha=0.14$, $R_M=1.57$) for dwarf ellipticals.\\ 
\begin{figure*} 
 \vspace{-0.5cm}
 \hspace{-1.5cm}
 \epsfig{file=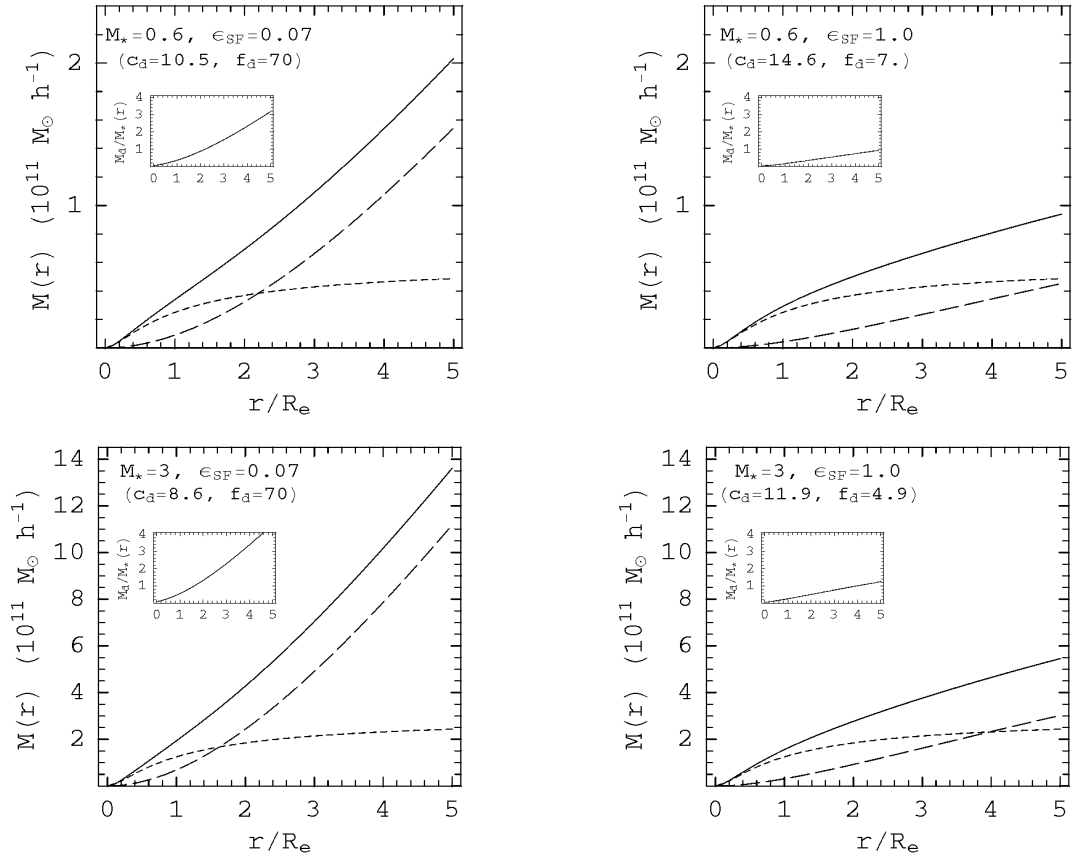,width=13.cm,height=18cm,angle=-90}
\caption{
Profiles of enclosed mass with radius.
Each panel shows a galaxy with parameters indicated.
The solid curve is the total mass, and the
dashed curves are the stellar and dark mass.
Insets show the dark-to-luminous mass fraction as a function of the radius. These grow almost linearly outside $1~R_{\rm e}$. Within $0.5~R_{\rm e}$ the dark matter fraction ranges between 10-20\%.
\label{mprof}} 
\end{figure*}  
\subsection{The dark component}
\label{cdm}
It is a basic prediction of CDM that visible galaxies are embedded in extended haloes of dark matter.
N-body simulations of hierarchical formation in the ``concordance'' $\Lambda$CDM cosmology 
($\Omega_{\rm m}=0.3, \Omega_{\Lambda}=0.7, \sigma_8=0.9$:
NFW; \citealt{bull01}; \citealt{wech02})
produce average dark halo mass profiles of the form
\begin{equation}
\rho_{\rm d}(r)=\frac{\rho_s}{(r/r_s)(1+r/r_s)^2} ,
\label{rhoNFW}
\end{equation}
where $r_s$ is the inner characteristic
length--scale, corresponding  to the radius where the  logarithmic
slope of the profile  is $-2$. Equation \ref{rhoNFW} can be written in terms of the 
total mass $M_{\rm d}$ at the virial radius $r_{\rm vir}$
and of the concentration parameter $c_{\rm d} \equiv r_{\rm vir}/r_s$:
\begin{equation}
\rho_{\rm d}(r)=\frac{M_{\rm d}}{4 \pi A(c_{\rm d})}\frac{1}{r(1+r)^2} ,
\label{rhoNFW2}
\end{equation}
where $M_{\rm d}=4 \pi \rho_s r_s^3 A(c_{\rm d})$ and 
\begin{equation}
A(x)=\ln (1+x)-\frac{x}{1+x} .
\end{equation}
The cumulative mass profile is then
\begin{equation}
M_{\rm d}(r) = M_{\rm d} \frac{A(r/r_s)}{A(c_{\rm d})} ,
\label{MNFW}
\end{equation}
where we have followed the notation of \cite{BSD03}.
In principle, NFW profiles are a two-parameter family of density distributions 
($\rho_s$ and $r_s$, or equivalently $M_{\rm d}$ and $c_{\rm d}$)\footnote{\cite{nav04} provide a more detailed analysis of the density profiles (see also \citealt{ML04b}), but using these profiles would change $M_{\rm d,out}$
and \dML\ by only $\sim$~10\%, which does not affect our conclusions in
this paper.}.
However, a key result of $\Lambda$CDM theory is that the halo concentration correlates on average 
strongly with the virial mass
(\citealt{bull01}; \citealt{wech02}), which
for a population of haloes at $z=0$ may be written as\footnote{
We have derived Eq.~\ref{cMvir} 
in our assumed $\sigma_8=0.9$ cosmology in the halo mass range $M_{\rm vir}=0.03-30\times 10^{12} M_{\odot}$
via the toy model code provided by J.~Bullock.}
\begin{equation}
c_{\rm d}(M_{\rm vir})\simeq 17.1 \left( \frac{M_{\rm vir}}{h^{-1} 10^{11} M_{\odot}}\right )^{-0.125},
\label{cMvir}
\end{equation}  
where the overdensity of the halo mass relative to the critical
density has been taken to be
$\Delta_{\rm vir}=101$ \citep{bull01}. In Eq. \ref{cMvir} $M_{\rm vir}$ 
includes the baryonic mass, $M_{\rm b}$, ($M_{\rm vir}=M_{\rm d}+M_{\rm b}$). Assuming a baryon fraction $f_{\rm b}=\Omega_{\mathrm{b}}/\Omega_{\mathrm{M}}=0.17$ (see next section), we have $M_{\rm vir} \sim 1.2 M_{\rm d}$ and Eq. \ref{cMvir} can be written explicitly for $M_{\rm d}$ as
\begin{equation}
c_{\rm d}(M_{\rm d})\simeq 16.7 \left( \frac{M_{\rm d}}{h^{-1} 10^{11} M_{\odot}}\right )^{-0.125}.
\end{equation}
Using this correlation,
NFW profiles can thus be considered as a one-parameter density distribution 
(characterised by $M_{\rm d}$ or $c_{\rm d}$).

The NFW profiles were produced in the framework of non-interacting (N-body) particles, but
real galaxy formation includes baryons, whose interactions could modify the dark matter mass
profiles significantly---especially in their centres.
Including these baryonic effects (cooling, star formation, heating, etc.) can be considered
the ``holy grail'' of galaxy formation theory, and there are currently efforts underway
in this direction \citep{meza03,sgp03,dh04,sds04,Wrig04,mm04}.
However, it seems that even the qualitative effects on the mass profiles are not yet known
with any certainty, so we are simply taking the NFW profiles as the first approximation to
the predictions of $\Lambda$CDM.
Some possible implications of baryonic effects are discussed in Section~\ref{imp}.
\subsection{The stellar mass fraction}
\label{frac}
From the previous sections, we see that early-type galaxy mass distributions in the 
$\Lambda$CDM framework can be approximated fairly well as a family of two parameters
(for instance, $M_{\rm d}$ and $M_*$). 
We would further like to parametrise the stellar-to-dark mass ratio $M_*/M_{\rm d}$.
Although dark matter could be stripped from the halo, and baryons could be lost or gained through
inflows and outflows, for want of more detailed knowledge we assume that the baryon fraction in
the galaxy halo is the same as the universal value: 
$f_{\mathrm{b}}=\Omega_{\mathrm{b}}/\Omega_{\mathrm{M}}$.
This latter quantity is well constrained from the measurements of the
cosmic microwave background, and
here we adopt $f_{\mathrm{b}}=0.17$ (Bennett et al. 2003). 
We relate this to the net star formation efficiency $\epsilon_{\rm SF}$
(including stellar mass loss) by the following:
\begin{equation}
f_{\mathrm{b}}^{-1}=\frac{M_{\rm vir}}{M_{\mathrm{b}}}=\frac{M_{\mathrm{b}}+M_{\rm d}}{M_{\mathrm{b}}}= 1+\frac{\epsilon_{\rm{SF}}~M_{\rm d}}{M_*} ,
\label{invfbar}
\end{equation}
and thus $\epsilon_{\rm SF} = 4.9 M_*/M_{\rm d}$.
While for physical insight, we will use $\epsilon_{\rm SF}$ as our key parameter in
this study, it should be remembered that this is a valid characteristic only if baryons are indeed conserved.
Thus we will also quote the more robust dark-to-luminous mass fraction parameter
$f_{\rm d} \equiv M_{\rm d}/M_*= 4.9/\epsilon_{\rm
SF}$.

There are no {\it a priori} constraints on $\epsilon_{\rm SF}$, although galaxy formation models use values in the range $\sim
0.2$--$1.0$ \citep{ben00}.
Its extreme upper limit (given the baryon conservation assumption)
is $\epsilon_{\rm SF}=1.0$ ($f_{\rm d}=4.9$), 
while its lowest limit can be taken from the universal average star formation efficiency
of $\epsilon_{\rm SF}=0.072$ ($f_{\rm d}=68$; \citealt{Fuk98}) and the assumption that early-type galaxies are at least as efficient as this,
since clusters and groups bring the average down significantly.
Studies matching mass and luminosity functions of virialised systems
have suggested $\epsilon_{\rm SF}\sim$~0.25
($f_{\rm d} \sim$~20) 
for typical galaxies \citep{mahu02,guse02}, while stellar masses in the SDSS survey \citep{pad04} implied $M_{\rm vir}/M_*=7-30$ and $f_{\rm d}\sim 6-25$ ($\epsilon_{\rm SF}\sim 0.2-0.8$). In the following we will consider conservatively $\epsilon_{\rm SF}\sim 0.07-1.0$ as our lower and upper limits respectively.
\begin{figure} 
\vspace{-0.5cm}
\hspace{-1cm}
\epsfig{file=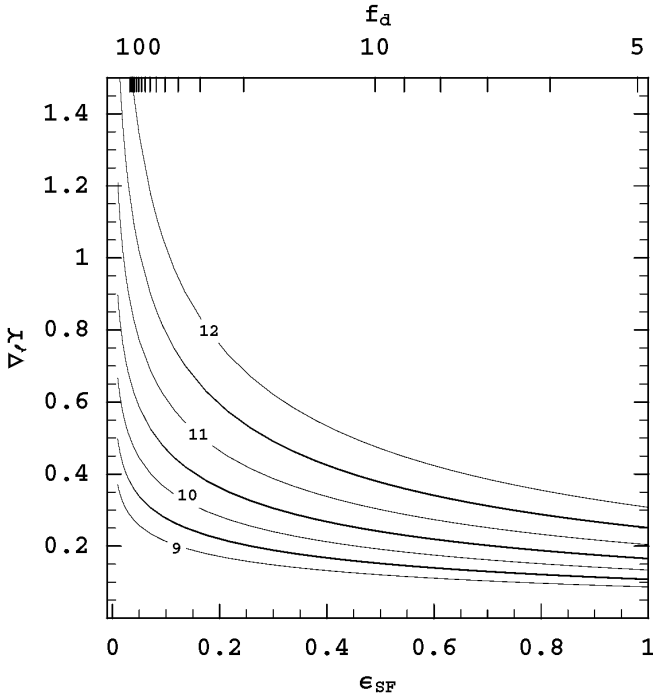,width=9.3cm,height=9.3cm,angle=0}
\caption{
Predictions for the \ML\ gradient versus star formation efficiency.
The curves are for stellar masses $\log (M_*/M_{\odot})$ of 
9.0 to 12.0 (marked in the plot), in increments of 0.5
(stellar luminosities $M_B$ of $-15.0$ to $-23.7$,
assuming $\Upsilon_*=6.5$).
The top axis shows the equivalent dark-to-luminous mass ratio
$f_{\rm d}=M_{\rm d}/M_*$.
\label{gradpred}} 
\end{figure} 
We thus construct a sequence of $\Lambda$CDM-based galaxy models characterised by $M_*$ and $\epsilon_{\rm SF}$.
In real galaxies, the scatter in the $R_{\rm e}$--$M_*$ and $c_{\rm d}$--$M_{\rm d}$ relations would produce
considerable variation about this sequence,
but for now we are considering the mean properties which should be apparent in
a large enough sample of galaxies.
In Fig.~\ref{mprof}, we show some example mass profiles,
where typical values for $M_*$ are assumed, and extreme cases of
$\epsilon_{\rm SF}=(0.07,1.0)$ are used.
Note that in all cases, the dark matter is the dominant mass component by 5~$R_{\rm e}$, 
indicating that early-type galaxies should generically show appreciable \ML\ gradients by this radius.
The fraction of the total mass in dark matter within 0.5~$R_{\rm e}$ is in these cases 10--20\% (increasing with $M_*$ and decreasing with $\epsilon_{\rm SF}$)\footnote{Testing
alternative parameter values in the $R_{\rm e}$--$M_*$ relation (Eq. \ref{remleq}) of $\alpha$=0.4--1.2 
does not change the conclusion
that the dark matter fraction at $R_{\rm in}$ is small.
}.
It is also evident in Fig.~\ref{mprof} that the \ML\ gradient is more strongly dependent on
$\epsilon_{\rm SF}$ than on $M_*$, which
arises because $R_{\rm e}/r_s$ scales weakly with mass (and thus galaxies with a given
$f_{\rm d}$ are a nearly homologous family).
In the same figure, inset panels show $M_{\rm d}(r)/M_{\rm *}(r)$;
this mass ratio does not significantly deviate from a linear growth curve
at $r \gsim 0.5 R_{\rm e}$, except for the case of very low $\epsilon_{\rm SF}$
(i.e., very dark matter dominated haloes) 
and thus the measurement locations in Eq.~\ref{grad} are generally not important
(see also next section).

\subsection{$M/L$ gradient predictions}
\label{pred} 
\begin{figure} 
\hspace{-1cm}
\epsfig{file=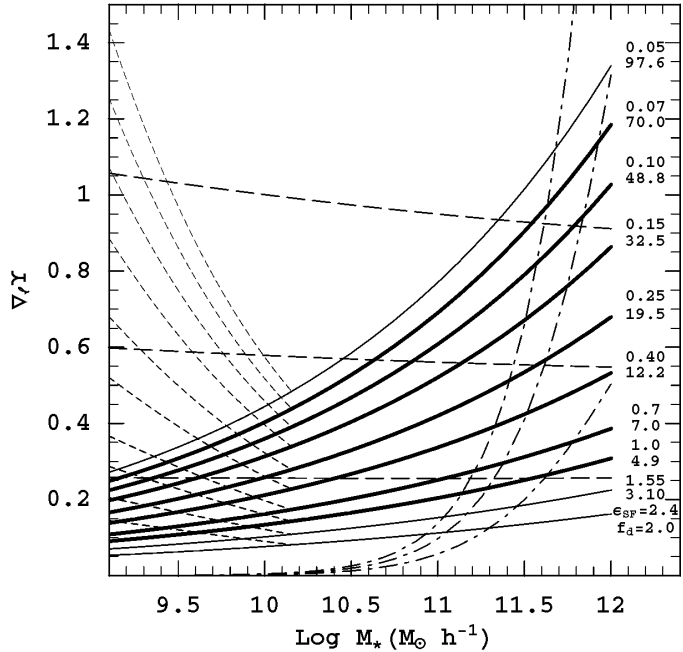,width=9cm,height=9cm,angle=-0}
\vspace{-0.15cm}
\caption{Predictions for the \ML\ gradient versus total stellar mass
from $\Lambda$CDM,
for different values of the star formation efficiency $\epsilon_{\rm SF}$.
The solid curves show solutions with the nominal $R_{\rm e}$--$M_*$ relation
of $(\alpha, R_M)=(0.56, 3.6)$;
they are labelled at the right side by the values of $\epsilon_{\rm SF}$
and the equivalent $f_{\rm d}$ (dark-to-luminous mass ratio).
Heavy curves are physically plausible solutions, and light curves are not. Short-dashed lines show the prediction for the shallower $R_{\rm e}$--$M_*$ relation for dwarf galaxies $(\alpha, R_M)=(0.14, 1.57)$: this is only indicative since it is based on an $R^{1/4}$ light distribution, which is not appropriate for dwarf systems.
Dot-dashed and long-dashed curves show the predictions assuming 
$(\alpha, R_M)=(0.4, 4.4)$ and $(1.2, 1.4)$, respectively;
solutions are plotted for $\epsilon_{\rm SF}= 0.07, 0.25, 1.0$.
\label{modfig}} 
\end{figure}  
We compute the model \ML\ gradients \dML\ (Eq.~\ref{grad}) 
with $r_{\mathrm{in}}=0.5 R_{\rm e}$ and $r_{\mathrm{out}}=4 R_{\rm e}$ 
representing typical values for the empirical measurements in Section~\ref{samp}.
In Fig.~\ref{gradpred}, we plot the predicted \dMLm\ versus $\epsilon_{\rm SF}$
for sample galaxy models with various values of $M_*$.
The gradient increases for decreasing $\epsilon_{\rm SF}$, 
since low-efficiency systems imply relatively more massive dark haloes.
In Fig.~\ref{modfig}, we plot the \dMLm\ versus galaxy mass $M_*$ for various values 
of $\epsilon_{\rm SF}$.
The gradient increases with $M_*$,
which can be qualitatively understood by the construction that
the scale radius of the stars varies more rapidly with mass ($R_{\rm e} \propto M_*^{0.56}$)
than the scale radius of the dark matter 
($r_s \propto r_{\rm vir} c_{\rm d}^{-1} \propto M_{\rm d}^{0.33} M_{\rm d}^{0.13} \propto M_{\rm d}^{0.46}$).
Thus the stellar body of a brighter galaxy encloses a larger fraction of its dark matter halo
than in a fainter galaxy, and appears more dark-matter dominated. This also means that in brighter galaxies, \dML\ is more
sensitive to $\epsilon_{\rm SF}$ than in fainter galaxies (see Fig. \ref{gradpred}).

Although these predictions appear uniquely determined for given $(\epsilon_{\rm SF}, M_*)$,
there is an $R_{\rm e}$--$M_*$ scaling relation implicitly assumed (Eq. \ref{remleq}),
which translates into assumed locations for $(r_{\rm in}, r_{\rm out})$.
We therefore check the effects on the model predictions of the 
$R_{\rm e}$--$M_*$ relation.
In Fig.~\ref{modfig} we show  the predicted \dML--$M_*$ curves for
$(\alpha, R_M) =$ $(0.4, 4.4)$ and $(1.2, 1.4)$,
in addition to the nominal values of $(0.56, 3.6)$.
The steep relation means a steeper trend for \dML:
for the more massive galaxies, $R_{\rm e}$ is larger for a given $M_*$, and thus the gradient
probes farther into the dark matter halo, producing higher values of \dML.
For the less massive galaxies, $R_{\rm e}$ is smaller, producing lower values of \dML.
With the shallow relation, $r_s$ scales more quickly than $R_{\rm e}$, and
the \dML\ behaviour goes in the opposite direction, actually declining slightly with $M_*$.
\begin{table*} 
\tiny 
\centering 
\hspace{-2.5cm}
\hspace{-1cm} 
\begin{minipage}{140mm}
\caption{
Catalogue of early-type galaxies with halo mass measurements.
}
\begin{tabular}{llrrrrlrrrrrrll} 
\hline 
\noalign{\smallskip} 
Galaxy&Type\footnote{Types and magnitudes from HyperLeda (http://www-obs.univ-lyon1.fr/hypercat/).}
&$D$\footnote{Distances from SBF measurements \citep{jt01}, rescaled from $h=0.74$ to $h=0.70$.
The exceptions are NGC 4486B---which is assigned the same distance as NGC 4486---and
NGC 1700, NGC 4464 and NGC 7626, for which the redshift-distance is used.
The distance uncertainties are taken from the SBF paper and included in
any uncertainties in $M_B$ and $M_*$, but are not included in the $\Upsilon$
uncertainties shown here since they cancel out when calculating \dML .
}
&$M_B$&$R_{\rm e}$\footnote{
For consistency,
$R_{\rm e}$ is taken from the dynamical reference paper.
} &
$a_4$ \footnote{ $a_4$ and $\gamma$ are taken from  \cite{BBF92,forb95,fab97,pel99,trag00,rav01}; and references therein.}
&$\gamma$&$r_{\rm in}$&$\Upsilon_{\rm in}$&$r_{\rm out}$&$\Upsilon_{\rm out}$
& $\Upsilon_{\rm SSP}$ & \dML\ & Data type\footnote{LS=long-slit integrated light stellar kinematics; PN=planetary nebulae; GC=globular clusters}; \\
&& (Mpc)& & (kpc) &&&($R_{\rm e}$)& ($\Upsilon_{B,\odot}$)&($R_{\rm e}$)&($\Upsilon_{B,\odot}$)&($\Upsilon_{B,\odot}$)&& $\Upsilon_{\rm in, out}$ Ref. \\ 
\hline 
NGC 221      &E3   &0.9  &-16.5 &0.2  &0     &0.5   &0.1 &$2.8 \pm0.2$  &5.6 &$4.8 \pm2.8$  & $3.0^{+0.9}_{-0.7}$ & $0.13 \pm0.18 $& LS,PN;MB01,V02\\
NGC 821      &E2   &25.5 &-20.6 &6.2  & 2.5  &0.64  &0.5 &$8.4 \pm0.4$  &4.8  &$13.1\pm3.9$  & $7.1^{+1.9}_{-2.0}$ & $0.13 \pm0.10  $&LS+PN;G03,R03\\
NGC 1316     &S0   &22.7 &-22.1 &12.0 & 0    &0     &0.6 &$3.2 \pm0.3$  &2.  &$5.7 \pm1.7$  & $3.9^{+0.8}_{-1.1}$ & $0.56 \pm0.39 $&LS+PN;A98\\
NGC 1379     &E0   &21.2 &-19.8 &2.5  &0.2   & -    &0.4 &$4.4  \pm0.4$  &2.5 &$4.4  \pm0.4$  & $7.4^{+2.5}_{-2.0}$ & $0.0  \pm0.07$&LS;MB01\\
NGC 1399\footnote{The PN and GC results from \cite{nap02} ($\ML \sim$~28--45 at 8.8 $R_{\rm e}$) imply \dML$\sim$~0.3--0.58 depending on assumptions about the halo equilibrium. We use the R04 result when computing correlations.}
     &E1/cD&21.1 &-21.4 &4.3  &0.1   & 0.07 &1.  &$8.3 \pm0.8$  &12  &$42  \pm13 $  & $10.4^{+3.3}_{-2.7}$& $0.37 \pm0.14 $&LS+GC;S00,R04\\
NGC 1700     &E4   &52   &-21.7 &3.5  & 0.4  &0.01  &0.5 &$4.  \pm0.4$  &4.6 &$7.8 \pm0.8$  & $3.3\pm0.4$         & $0.23 \pm0.06$&LS;S99\\
NGC 2434     &E1   &22.8 &-20.5 &2.7  & 0    &0.7   &0.5 &$8.2 \pm0.3$  &2.6 &$15.8\pm3.4$  &--                   & $0.44 \pm0.20  $&LS;K00\\
NGC 3115     &E6/S0&10.2 &-20.2 &2.8  &1.0   &0.78  &0.8 &$6   \pm0.6$  &3.3 &$10  \pm2 $  &--                   & $0.27 \pm0.14 $&LS;C93,E99\\
NGC 3379     &E1   &11.2 &-20.1 &2.0  &0.2   &0.18  &0.5 &$5.9 \pm0.4$  &7.9 &$8.7 \pm1.1$  & $8.4^{+2.8}_{-2.4}$ & $0.06 \pm0.03 $&LS+PN;G03,R03\\
NGC 3384     &E5/S0&12.3 &-19.8 &1.5  &1.0   &0.6   &0.5 &$4.1 \pm0.4$  &5.8 &$7.4 \pm2.2$  & $3.2\pm0.6$         & $0.15 \pm0.1  $&LS,PN;T95,B96\\
NGC 4406     &E3   &18.1 &-21.6 &8.9  &-0.7  &0.08  &0.3 &$6.4 \pm0.7$  &2.4 &$11.4\pm3.4$  & $8.1^{+2.4}_{-2.2}$ & $0.37 \pm0.23 $ &LS,PN;M91,A96\\
NGC 4464     &S0/a &15.3 &-18.7 &0.45 &0.5   &0.88  &0.5 &$7.9 \pm0.8$  &2.8 &$8.7 \pm2.6$  & $4.6^{+1.7}_{-0.7}$ & $0.04 \pm0.15 $&LS;MB01\\
NGC 4472     &E2   &17.2 &-22.  &8.7  &-0.3  &0.04  &0.5 &$8.  \pm0.3$  &4.5 &$28.5\pm8.6$  & $8.4^{+2.2}_{-2.3}$ & $0.64 \pm0.27 $ &LS,GC;K00,C03\\
NGC 4486     &E3/cD&17.  &-21.7 &7.8  &0     &0.2   &0.5 &$5.3 \pm0.4$  &4.8 &$30. \pm4.5$  &--                   & $1.1  \pm0.2  $&LS+GC;RK01\\
NGC 4486B    &E1   &17.  &-16.8 &0.26 &0.8   &0.14  &0.2 &$8.1 \pm0.6$  &2.8 &$10.3\pm2.2$  &--                   & $0.1  \pm0.11 $&LS;K00\\
NGC 4494     &E1   &18   &-20.7 &4.3  &0.3   &0.6   &0.5 &$3.9 \pm0.4$  &3.9 &$5.5 \pm1.7$  &--                   & $0.12 \pm0.13 $ &LS,PN;K00,R03\\
NGC 4697     &E3   &12.4 &-20.2 &5.7  &1.4   &0.74  &0.4 &$9.3 \pm0.4$  &3.4 &$9.3 \pm2.8$  & $6.7^{+2.3}_{-1.7}$ & $0.0   \pm0.1  $&LS,PN;M01\\
NGC 5128     &S0   &4.4  &-21.  &6.6  &-0.5  &0.15  &0.3 &$2.8 \pm0.3$  &10  &$10 \pm3$  &--                   & $0.27 \pm0.07$ &PN;P04\\
NGC 5846     &E1   &26.3 &-20.4 &8.0  &0     &$<$0.2&0.5 &$9.  \pm1. $  &3.2 &$20  \pm6 $  & $11.1^{+3.7}_{-3.2}$& $0.45 \pm0.25 $&LS;MB01\\
NGC 6703     &E1   &28.2 &-20.3 &4.1  &0     &-     &0.5 &$5.3 \pm0.3$  &2.6 &$7.6 \pm2.3$  & $4.7\pm1.0$         & $0.21 \pm0.21 $&LS;K00\\
NGC 7626     &E2   &46   &-21.4 &8.5  &0.2   &$<$0.2&0.5 &$10. \pm0.6$  &2.1 &$18.3\pm3.6$  & $10.8^{+3.6}_{-3.1}$& $0.52 \pm0.23 $&LS;K00\\
         
\hline 
\\
\noalign{\smallskip} 
\end{tabular} 
\label{datatable} 
\end{minipage}
\end{table*}   

Note that the mass of any diffuse gas [i.e., $M_{\rm g}=M_* (1-\epsilon_{\rm SF})$]
should add to the \ML\ gradient as a baryonic dark matter component.
However, assuming that this gas is distributed diffusely out to the virial radius (following the CDM density profile),
its contribution to \dML\ is no more than $\sim 0.02$.
We have also assumed that $\Upsilon_*$ is constant with radius, even if observed colour gradients of galaxies suggest a slightly decreasing with radius.
Although it would be ideal to make our measurements in the $K$-band rather than 
the $B$-band in order to minimise this effect, we estimate the contribution
to \dML\ in our current calculations to be only $\sim -0.02$.
We also test the sensitivity of \dML\ to its measurement locations $(r_{\rm in}, r_{\rm out}$);
these are important only for very dark-matter dominated systems, such that
a value of \dML~$\sim 1.0$ changes by $\sim0.1$ if $\Delta r_{\rm out}\sim\pm2 R_{\rm e}$.\\
Our model galaxy constructions based on $\Lambda$CDM theory thus make
specific predictions for mass-to-light ratio gradients. The quantity
\dML\ increases with $M_*$ unless $\epsilon_{\rm SF}$ systematically
increases with $M_*$, or the scaling parameter $\alpha$ is very
high.  If dwarf ellipticals do indeed follow a shallower scaling
relation $\alpha$, then \dML\ should increase again at low $M_*$ (c.f. \citealt{deksil86}), with 
a minimum occurring at $\log M_*/h^{-1} M_{\odot} \sim 10.2$.

\section{Observational results}
\label{obs}
We describe our compilation of \ML\ results for 
early-type galaxies in Section~\ref{samp}.
We study the $R_{\rm e}$--$M_*$ relation in Section~\ref{reff},
and search for correlations between parameters in Section~\ref{corr}.
Hereafter we assume $H_0=70$~km~s$^{-1}$~Mpc$^{-1}$.
\subsection{Galaxy sample}\label{samp}
We assemble from the literature the dynamical results on all early-type galaxies with
mass measurements at a radius $r_{\rm out} \gsim 2 R_{\rm e}$.
These exclude galaxies with masses derived through X-ray analyses, strong gravitational lensing and gas disks and rings, as these techniques may be prone to strong selection effects. 
For instance, X-ray emission occurs mostly in giant elliptical galaxies, so these results are restricted to the brightest, most massive systems---mostly group or cluster dominant ellipticals (see \citealt{OsPon04b} for a discussion). 
Gravitational lensing also suffers a selection effect for large masses, 
since the lensing cross-section is a power-law of the lens mass.
Furthermore, the systems studied by lensing are generally located at large redshifts $0.1<z<1$ \citep{TrKoo04}: here 
it appears that
low mass galaxies are still forming stars while only very massive galaxies are already in place \citep{jim04}. This makes the selection effect even worse if we want to compare results with local galaxies. 
Extended HI rings (which are rare occurrences)
are mostly located in S0 systems. Furthermore, in most of the cases, these systems are gas rich ($M_{\rm HI}/L_{\rm B,\odot}\gsim 0.2$) which 
suggests 
that these galaxies are the result of a particular formation history (merging remnants of gas rich disks, \citealt{Oost02}) and are not representative of the ellipticals on the whole.
Thus, none of these methods is really suitable for an unbiased survey of the properties of dark matter haloes around elliptical galaxies. 

The techniques remaining include integrated stellar light, PNe and GCs;
there is no obvious reason for any of these to produce a particular selection effect on total mass (rather than on luminosity).
The results could in principle be skewed by dust effects when
relying on integrated stellar light kinematics \citep{baes},
but the halo masses of most of the sample galaxies are derived from
discrete tracers (PNe and GCs),
which are not affected by this issue.
There are 21 galaxies in the sample (16 ellipticals, 3 lenticulars and 2 of a transitional nature);
their properties are summarised in Table \ref{datatable},
including $a_4$ (isophote shape parameter) and $\gamma$ (inner surface brightness slope).
Note that this sample includes two cD galaxies, whose halo masses may very well be dominated
by the galaxy cluster in which they reside. However, this may also be true of group-dominant galaxies in general, and
at this point we do not wish to make such {\it a priori} distinctions; we will check any density environment dependence later on.

Some of the dynamical mass-to-light ratios 
($\Upsilon_{\rm in}, \Upsilon_{\rm out}$---provided in the $B$-band for uniformity)
are derived from highly flexible modelling techniques which fit orbital distributions
to the detailed line-of-sight velocity distributions \citep{K00,kg03,ral03}.
For these we take the uncertainties in $\Upsilon$ as given in the papers.
In other cases, Jeans (and similar) models have been used to fit binned velocity dispersion profiles;
in such models, orbital anisotropy is a major source of systematic uncertainty.
Although not many detailed dynamical studies have extended beyond 2~$R_{\rm e}$,
so far no cases of extreme anisotropy have been detected (see e.g., \citealt{gerhard01}).
We therefore assume that the anisotropy parameter $\beta_{\rm a}$ lies in the range $\pm0.3$
($\beta_{\rm a}=0$ is isotropic), and estimate uncertainties in 
$\Upsilon_{\rm in}$ of $\pm10$\% and in $\Upsilon_{\rm out}$ of $\pm30$\%.
\begin{figure} 
\hspace{-0.5cm}
 \epsfig{file=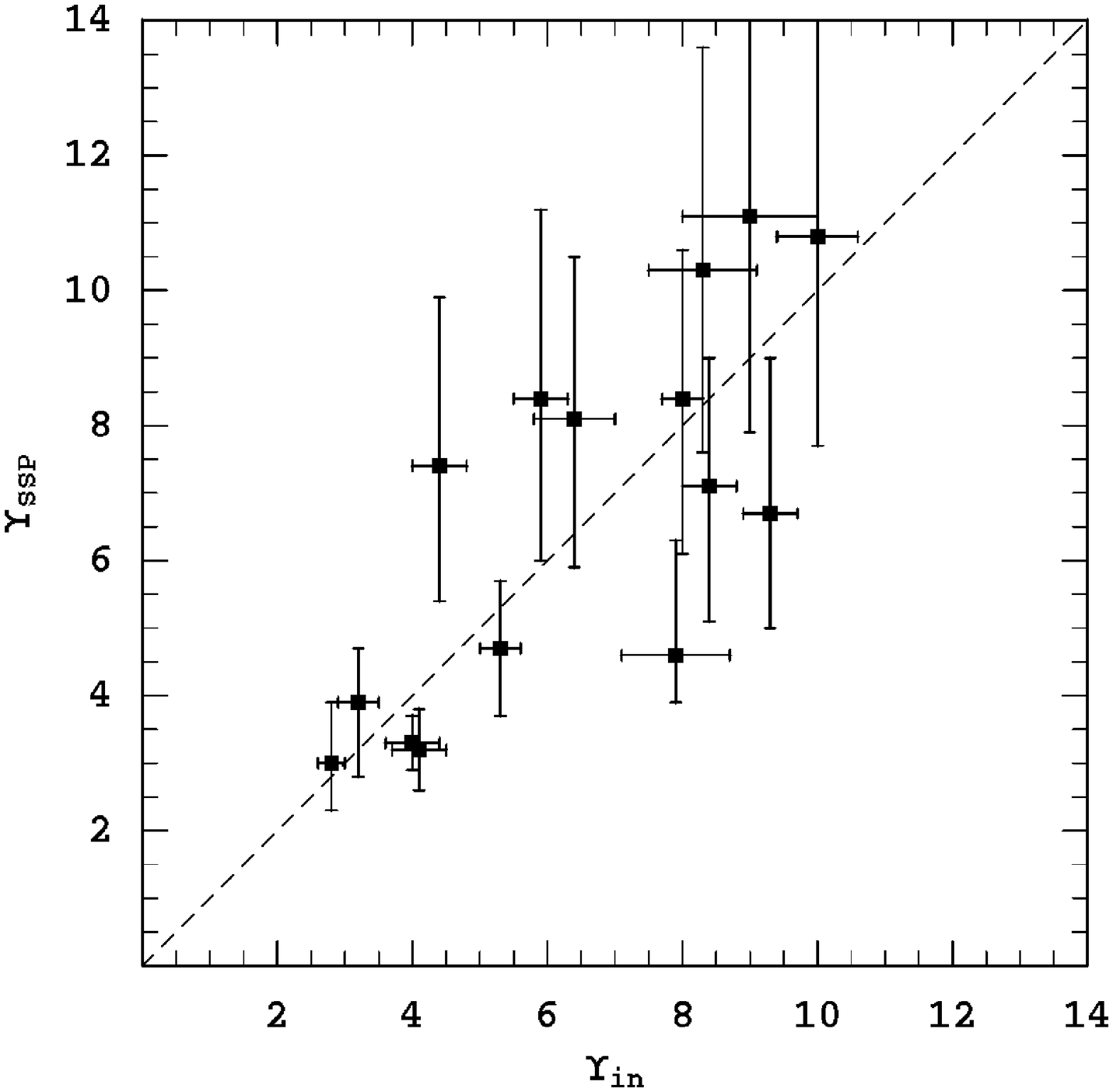,width=9.cm,height=9.cm,angle=0}
\vspace{-0.5cm}
\caption{
Estimates of stellar \ML\ from stellar population synthesis modelling ($\Upsilon_{\rm SSP}$)
compared to dynamical modelling ($\Upsilon_{\rm in}$),
in units of $\Upsilon_{B,\odot}$.
The dashed line shows $\Upsilon_{\rm in}=\Upsilon_{\rm SSP}$.
\label{mlml}}
\end{figure}
  
To find the empirical gradient \dMLo, we also need to know $\Upsilon_*$, 
which may be estimated in two completely independent ways.
We can use spectroscopic information for the luminous body combined with (single-burst) stellar
population synthesis models: $\Upsilon_* = \Upsilon_{\rm SSP}$.
This is in principle the most direct way to measure the stellar mass only, 
but currently such models are susceptible to considerable systematic uncertainties.
The second way is to use the dynamical estimate: $\Upsilon_* = \Upsilon_{\rm in}$;
there will be some dark matter contribution to $\Upsilon_{\rm in}$, 
so this approach will provide a lower limit on \dMLo.
The advantage of this approach is that many systematic uncertainties (such as distance)
will cancel out in Eq.~\ref{grad}; 
the empirical gradient is thus:
\begin{equation}
\dMLo\ = \frac{R_{\rm e}}{\Delta r}\left(\frac{\Upsilon_{\rm out}}{\Upsilon_{\rm in}}-1\right).
\end{equation}
\label{gradeqn}

We first check the reliability of these two approaches by comparing their results in
a common galaxy subsample.
The mean stellar age and metallicity are taken from \cite{TF02},
or if unavailable, from \cite{trag00} or from \cite{hk01};
note $r_{\rm in}=R_{\rm e}/8$ generally.
We then obtain the expected $B$-band stellar mass-to-light ratio, $\Upsilon_{\rm SSP}$, 
using stellar population synthesis models from \cite{W94}.
After examining galaxies in common in order to tie the results onto a uniform scale,
we increase the $\Upsilon_{\rm SSP}$ values from
\cite{trag00} by 15\% and decrease those from \cite{hk01}  by 8\%.
The results for the 15 galaxies with available age and metallicity are listed in Table~\ref{datatable}.
\begin{figure} 
\vspace{+0.2cm}
\hspace{-0.4cm}
\epsfig{file=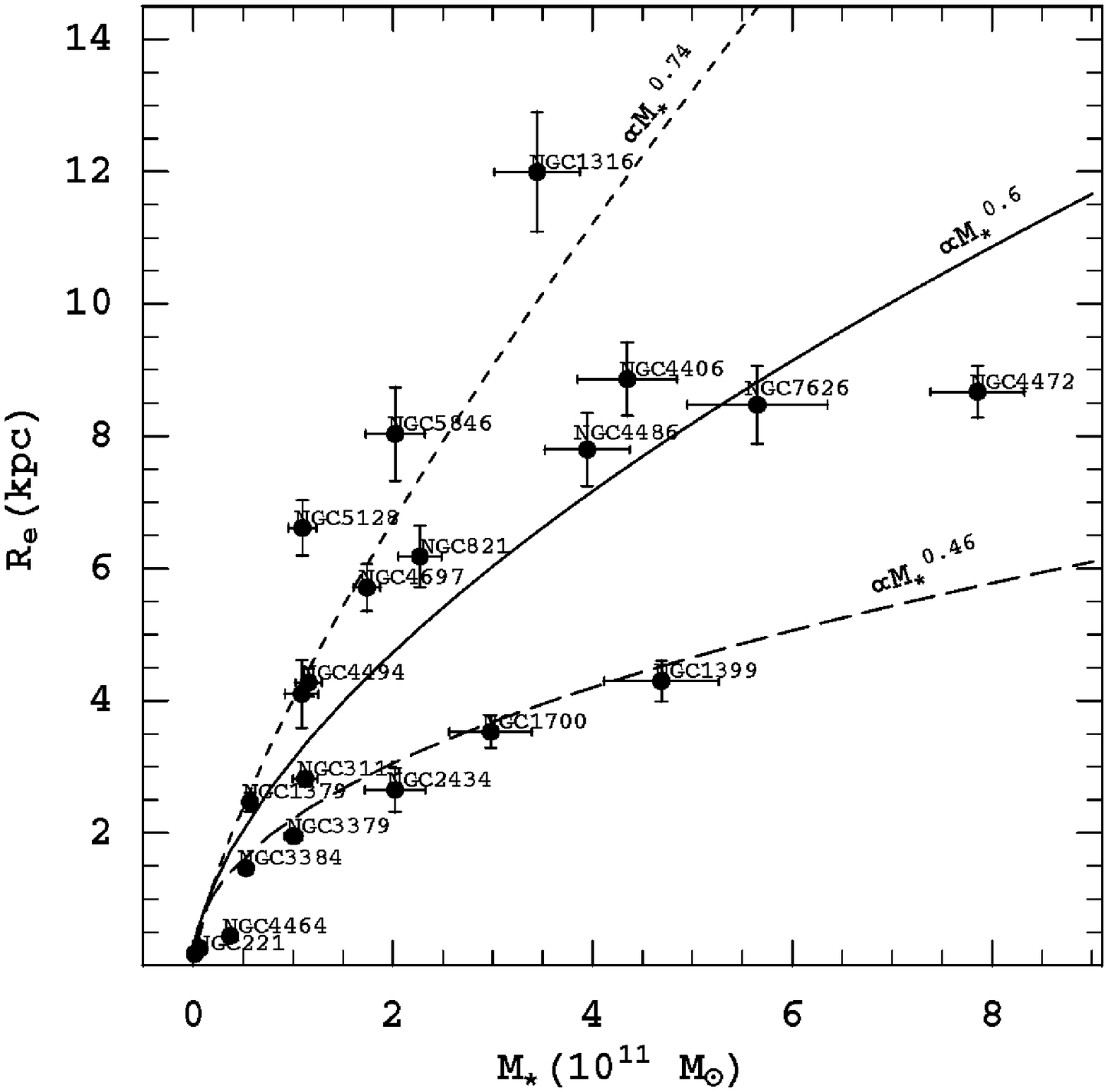,width=8.8cm,height=8.8cm,angle=0}
\vspace{-0.5cm}
\caption{
Relation between total luminous mass and effective radius.
The solid line is the best fit to the data $(\alpha=0.6, R_M=3.2$)
and the dotted and dashed lines show the $\pm$~1~$\sigma$ uncertainties of the fit:
$(\alpha=0.74, R_M=4.0$) and $(\alpha=0.46, R_M=2.4$), respectively.
\label{fig1}}
\end{figure} 
We compare $\Upsilon_{\rm SSP}$ to the innermost dynamical estimates $\Upsilon_{\rm in}$ in Fig.~\ref{mlml},
and find them to be remarkably consistent given all the potential sources of error
which have not been included in the uncertainties
(e.g., distance, IMF, heterogeneous $r_{\rm in}/R_{\rm e}$).
This consistency indicates that if some fraction of $\Upsilon_{\rm in}$ comes from dark matter,
then $\Upsilon_{\rm SSP}$ has been systematically overestimated by approximately the same amount.
We will thus assume $\Upsilon_*=\Upsilon_{\rm in}$ for all our ensuing analyses, 
also examining in due course the potential effects of a non-negligible dark matter contribution.

A best fit of $\Upsilon_*$ to galaxy luminosity
and mass shows a weak correlation:
$\Upsilon_{B,*} \propto L_B^{0.07\pm0.06}$,
$\Upsilon_{B,*} \propto M_*^{0.10\pm0.06}$.
This is weaker than some of the trends found in the literature
(see Section~\ref{lum}), but
\cite{kauffmann03} and \cite{pad04}
also find $\Upsilon_*$ to be roughly constant.
The models of the latter two imply $\Upsilon_* \sim$~5 in the $B$-band,
while we obtain $\sim$~7 typically, lending credence to the possibility
that we are underestimating the amount of dark matter in the central
parts of our galaxy sample.

\subsection{Mass-radius relation}
\label{reff}
As seen in Section 2, to derive the
\ML\ gradient predictions, we needed a mean relation between $R_e$ and
$M_*$ (Eq. \ref{remleq}), where we adopted values for the parameters
$(\alpha,R_M)$ based on literature results. For direct comparison to our
empirical \ML\ gradient results, we must check that our galaxy sample is
consistent with this relation.  We begin by fitting a mean $R_e$-$M_*$
relation to our sample (as shown in Fig. \ref{fig1}), where $M_*$ is obtained by multiplying
$\Upsilon_*$ and the total luminosity $L_B$.
We find $(\alpha, R_M)=(0.50
\pm 0.13, 3.6 \pm 0.8)$, which is consistent with $(0.56,4.2$; $h=0.7$)
from \cite{shen03}.  Since this fit is sensitive to sparse sampling
effects in our data set, we instead wish to use an exponent $\alpha$ based
on the much larger SDSS galaxy sample, with the normalisation $R_M$ set by
our data because of systematic differences in measuring masses and sizes.  
Assuming a typical $\alpha=0.6$ (see Section \ref{lum}), we find $R_M=3.1\pm0.8$; for comparison $R_M=4.2$
from \cite{shen03} \footnote{This is consistent with their
$\Upsilon_*$ values
(obtained via population synthesis modelling; \citealt{kauffmann03})
being systematically lower than ours.},
and the \cite{pahre98} results (where also $\alpha\sim 0.6$) imply
$R_M=2.6$ if $(B-K)=4.1$ and $\Upsilon_*=7$.
Thus, our $R_{\rm e}$--$M_*$ relation
is intermediate between typical literature values.
\begin{figure*} 
\epsfig{file=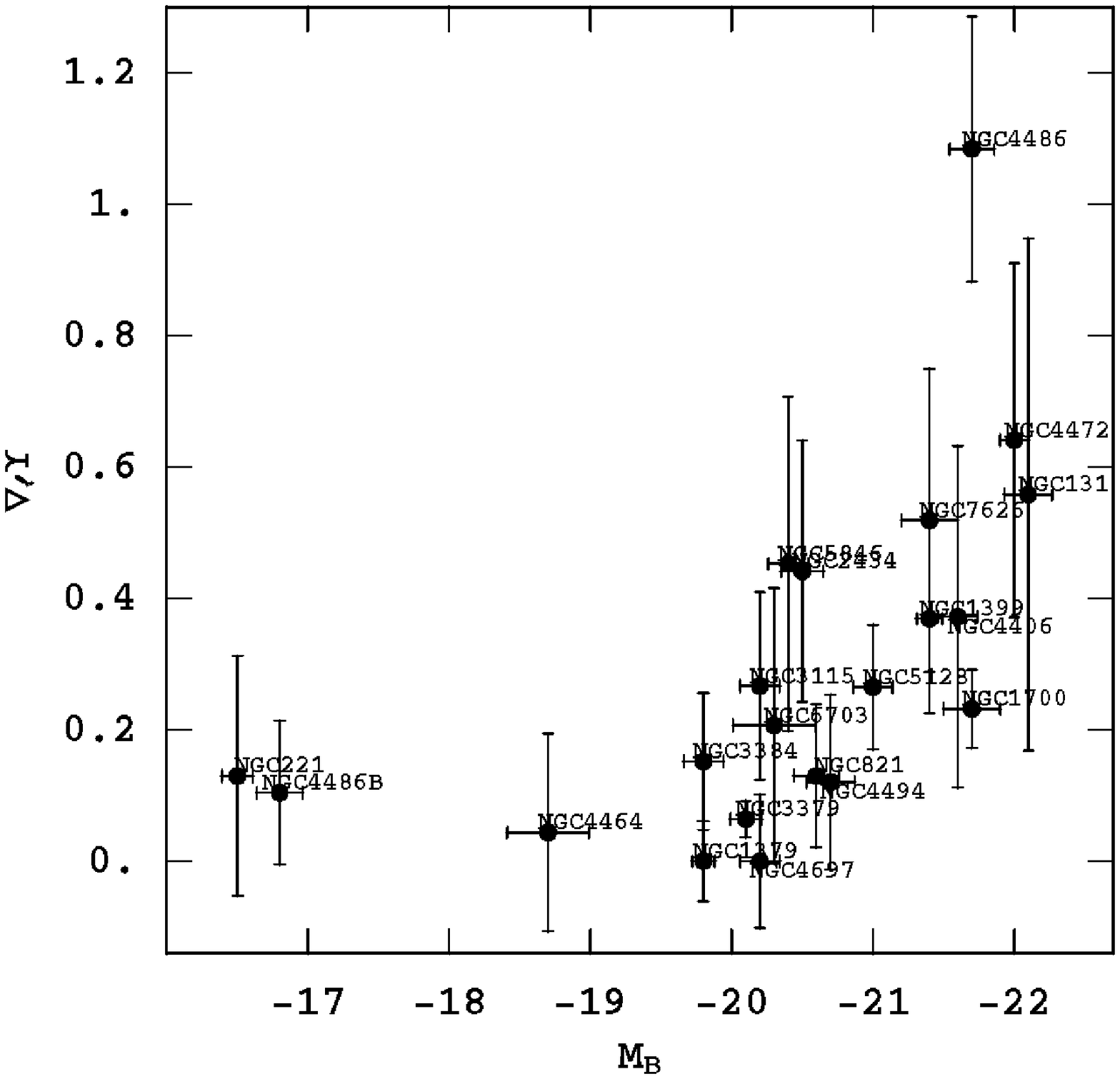,width=8.75cm,height=8.75cm,angle=-0}
\hspace{-0cm} 
\epsfig{file=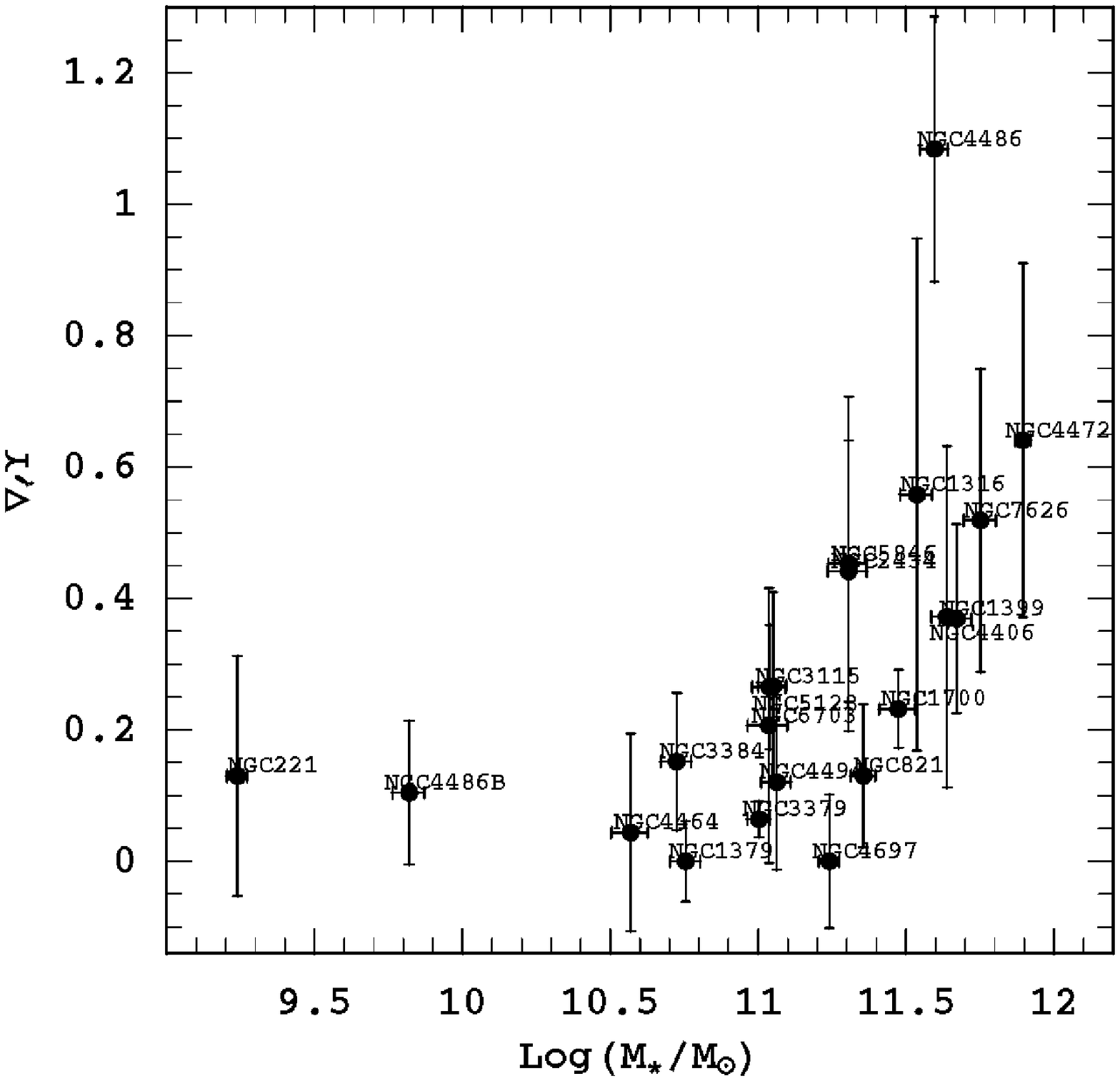,width=8.75cm,height=8.75cm,angle=-0}
\hspace{-3.4cm}
 \epsfig{file=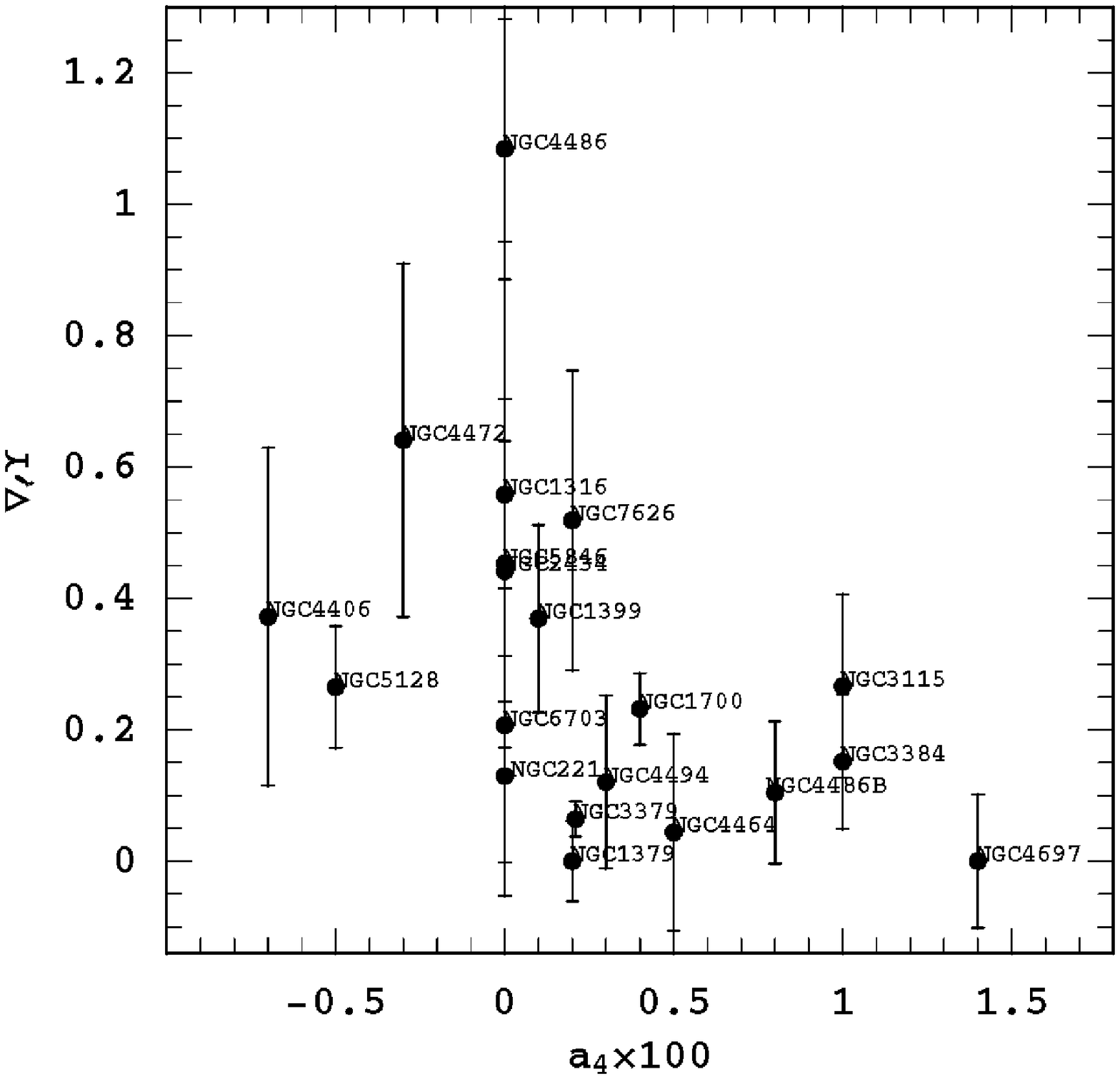,width=8.75cm,height=8.75cm,angle=-0}
 \hspace{-0cm}
 \epsfig{file=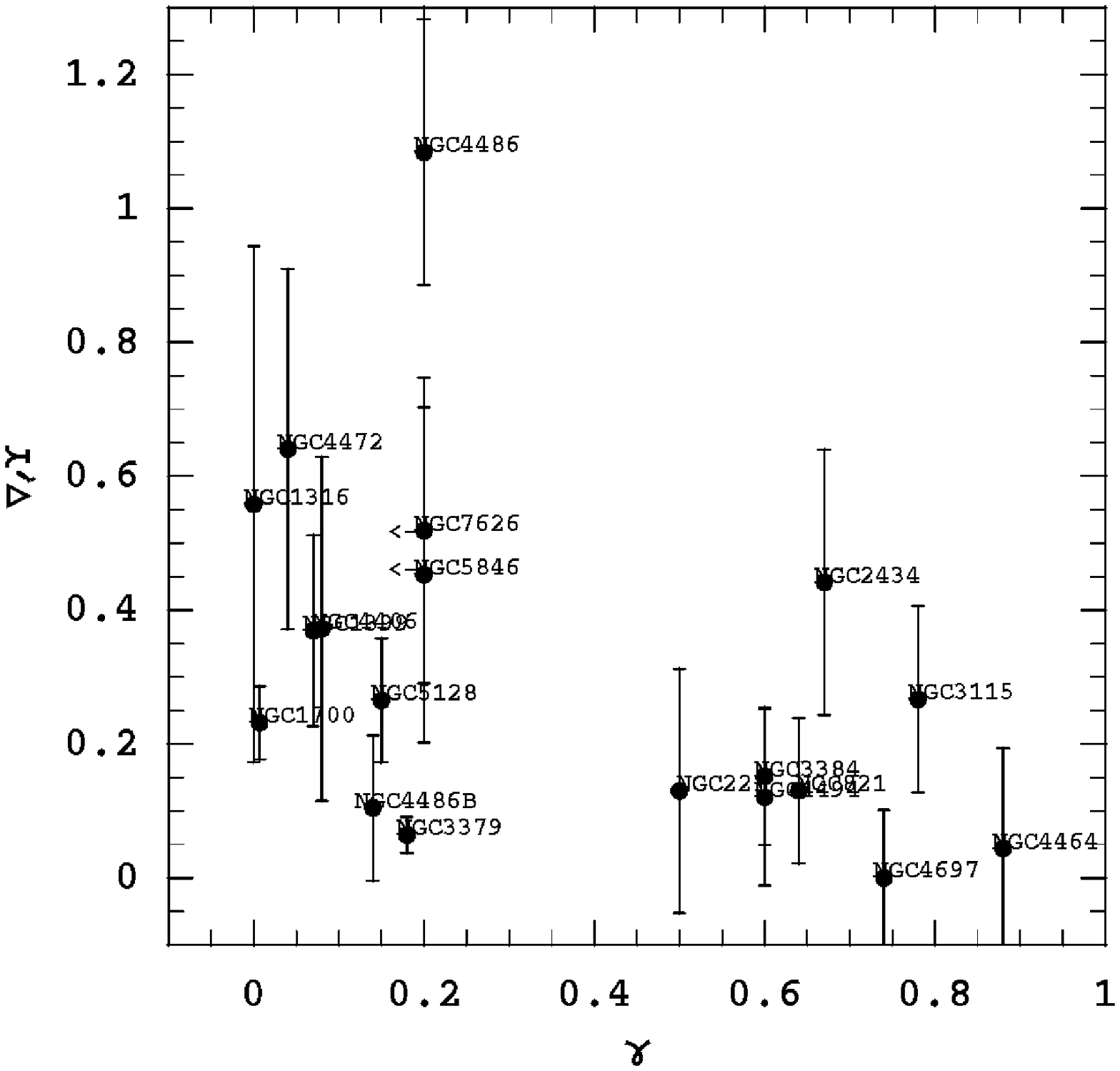,width=8.75cm,height=8.75cm,angle=-0}
\caption{Empirical $M/L$ gradients versus galaxy parameters: 
the total luminosity $M_{\mathrm{B}}$ (upper left), the
total stellar mass $M_*$ (upper right), the
isophotal shape $a_4$ (lower left), and the central surface brightness
slope $\gamma$ (lower right). Error bars are 1-$\sigma$ level.
\label{fig0}}
\end{figure*}  
The uncertainties in this relation
are important to consider when making predictions for \dML.
In particular,
the exponent $\alpha$ is not yet known conclusively and
may systematically vary with $M_*$;
we have seen in Section~\ref{pred} how changes in $\alpha$ would affect the model
predictions for \dML.
We have also seen that,
even if $\alpha$ is known accurately, there is a large scatter in 
real galaxies around the
mean $R_{\rm e}$--$M_*$ relation
(which is after all only a projection of the fundamental plane).
So when we later come to \ML\ results in Fig.~\ref{fig2}, these should not be over-interpreted for any single galaxy;
rather, the mean trends in a large sample of galaxies are meaningful.
We can quantify the effects of the scatter by finding an ``envelope'' for the
$R_{\rm e}$--$M_*$ relation, enclosing 68\% of the data points.
The upper and lower envelopes have $(\alpha, R_M)=(0.74,4.0)$ and $(0.46, 2.4)$, respectively
(see Fig.~\ref{fig1});
these values are well within the plausible bounds on $\alpha$ of 0.4--1.2 discussed in Section~\ref{lum}.
Thus when we compare to theoretical
models in Section 5, we can safely use $(\alpha=0.6, R_M=3.1)$, based on
larger galaxy surveys, as representative of our sample; and we can check
the effects of allowing $(\alpha, R_M)$ to vary within plausible limits.

\subsection{Correlations}
\label{corr}

Before comparing the empirical results on the \ML\ gradient to theoretical predictions, 
we would like to see if there are any correlations in the data
(see earlier work in \citealt{cap03,nap04}).
The broadest property to test against is the total galaxy magnitude $M_B$.
Plotting \dMLo\ against $M_B$ in Fig.~\ref{fig0}, 
we see that fainter galaxies have shallow gradients
($\dMLo \sim 0.0$--0.1 for $M_B \lsim -20$)
while the brighter galaxies have a wide range of gradients
($\dMLo \sim 0.1$--0.6). The same trend is found when considering galaxy mass $M_*$ (as estimated using $\Upsilon_*=\Upsilon_{\rm in}$) rather than galaxy luminosity, with $\log M_*\sim 11.2$ marking the transition between shallow and steep gradients. 
We also see evidence in Fig.~\ref{fig0} for disky galaxies ($a_4 \gsim 0.2$) having
shallow gradients while boxy ones have a wide range,
and for galaxies with a steep central cusp ($\gamma \gsim 0.2$) having shallower gradients
than those with a shallow cusp.\\
These last two trends are not surprising given the first, since it is well-established that
galaxy properties such as $a_4$ and $\gamma$ correlate strongly with luminosity \citep{nieto,cap92,fab97}.
\begin{table}
\flushleft 
\centering 
\begin{minipage}{140mm}
\caption{Spearman correlation test }
\begin{tabular}{llll} 
\hline 
\noalign{\smallskip} 
Correlation & $N$ & $r_{\rm{spear}}$  & confidence level\\ 
\hline 
\dML\ -- $M_B$  & 21 & -0.58 & 99.9\%\\
\dML\ -- $a_4$  & 21 & -0.42 & 98\%\\
\dML\ -- $\gamma$ & 19 & -0.39 & 97\%\\
\dML\ -- $M_*$ & 21 & 0.58 & 99.9\%\\
\dML\ -- $\rho_{\rm env}$ & 18 & 0.10 & 78\%\\
\hline 
\noalign{\smallskip} 
\end{tabular}
\end{minipage}
\label{SpSt} 
\end{table}  
The strengths of the correlations as measured by the Spearman rank statistic $r_{\rm spear}$ \citep{numrec} are given in Table 2, where
$N$ is the number of galaxies.
To estimate the statistical significance of these correlations, 
we performed, for each structural parameter, 10\,000 random experiments where \dML\ have been randomly extracted for each galaxy 
according to a Gaussian measurement distribution.
The confidence level of the correlations being real is then given in the table.\\
The correlations apparent by eye are all significant at the more than 97\% level,
and the ones with $M_B$ and $M_*$ are the strongest. 
Note that these correlation results are independent of any distance uncertainties
(which cancel out in the method used for estimating \dMLo).
Nor are they changed if $\Upsilon_*$ is systematically wrong in the sense that there is a large,
constant fraction of dark matter in the galaxy centres:
this would change \dMLo\ equally for all galaxies and would not affect their ordering.
A possible effect is if $\Upsilon_*=\Upsilon_{\rm in}$ is wrong by an amount which varies systematically with $M_B$.  

However, in our basic $\Lambda$CDM picture, $\Upsilon_*$ is likely to be systematically 
too {\it high} for the brightest galaxies (see Section~\ref{frac}), leading to \dMLo\ estimates which are too
{\it low}; thus, the ``corrected'' values of \dMLo\ would result in an even stronger correlation.
Even if $\Upsilon_*$ were wrong in the opposite sense 
(such that $\Upsilon_*/\Upsilon_{\rm in}$ is actually much lower for low-$L$ ellipticals),
we estimate that dark matter mass
fractions of $\sim50$\% would be needed to explain the correlation---much larger than the
fractions typically estimated from other studies.\\
As a final source of systematic effects in the \ML\ trends, we have checked any dependence of \dMLo\ with the (density) environment. 
We used for this purpose the galaxy density estimates $\rho_{\rm env}$ (number of galaxies per Mpc$^3$) from \citet{tull88} and compared with our \dMLo. We did not find any significant correlation with \ML\ gradients: the Spearman rank statistics gives a 78\% significance for a positive correlation. The steepest gradients (NGC 4486 and NGC 4472) are indeed the ones related to the densest environment (but note that their close companions NGC 4486B and NGC 4464 have low gradients). These are the only apparent cases where the environmental effect could have enhanced the gradients because of a cluster dark matter background or the effect of a close encounter (see \citealt{nap02} for NGC 1399).\\
We therefore conclude that there is a empirical trend for bright, boxy shallow-core 
early-type galaxies to appear more dark-matter dominated than faint, disky steep-core galaxies.
This result is irrespective of the type of underlying dark matter distribution and environment, but is
especially strong if it resides in basic $\Lambda$CDM haloes.\\
\begin{figure} 
\vspace{-0.3cm}
\hspace{-0.5cm}
\epsfig{file=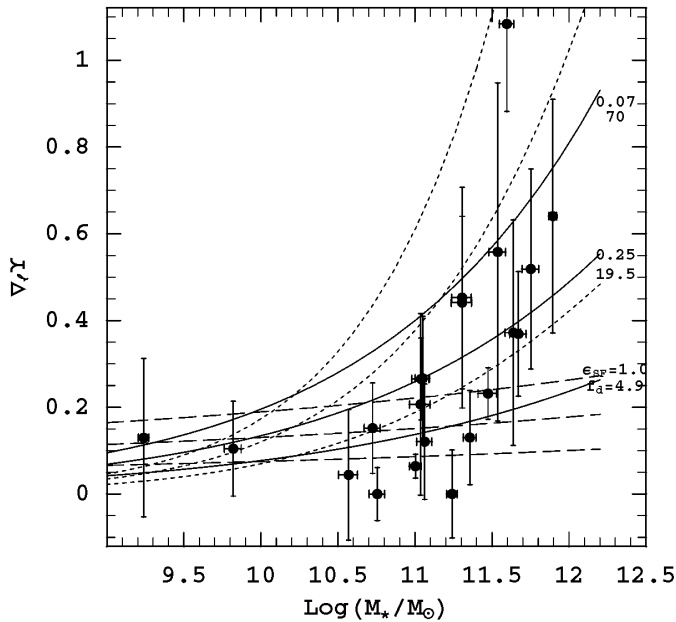,width=9.cm,height=9.cm,angle=-0}
\vspace{-0.65cm}
\caption{
\ML\ gradients versus total luminous mass. 
Points with error bars show galaxy data with uncertainties.
Labelled curves are our predictions from $\Lambda$CDM for different 
values of $\epsilon_{\rm SF}$ and $f_{\mathrm{d}}$: solid curves are models for the best-fit $R_{\rm e}$--$M_{\rm *}$ relation ($\alpha=0.6$, $R_{\rm M}=3.6$), short dashed and long-dashed curves are models for the 68\% envelopes as in Fig. \ref{fig1}. \label{fig2}} 
\end{figure}  

\section{Implications of predictions vs. observations}\label{imp}
In Section~\ref{nomres},
we show the results on halo masses and implied $\epsilon_{\rm SF}$ 
using the \dMLo\ data from Section~\ref{obs}
in conjunction with the model predictions of Section~\ref{pred}.
Finding many of the halo masses to be problematically low in the context of $\Lambda$CDM,
we consider possible alternative explanations in Section~\ref{poss}. In Section \ref{otech} we briefly discuss the results based on other observation techniques (X-rays, gravitational lensing and HI rings) and we lastly give some indication of the $\epsilon_{\rm SF}$ inferred in a few systems from the very low-mass regime in Section \ref{lowmass}.
\begin{figure} 
\vspace{-0.4cm}
\hspace{-0.8cm}
\epsfig{file=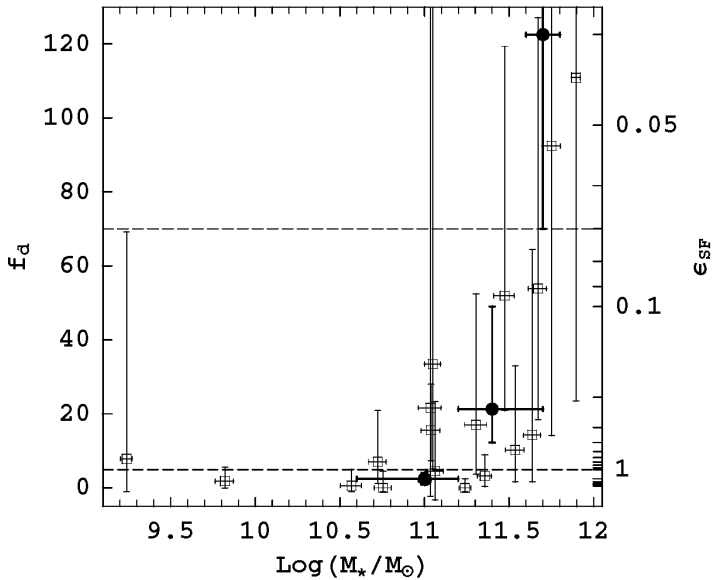,width=9.5cm,height=9.5cm,angle=-0}
\vspace{-1cm}
\caption{Efficiency of star formation versus total luminous mass,
for early-type galaxy data fitted to $\Lambda$CDM models (open squares).
Plausible upper and lower bounds on these parameters are shown as dotted lines. Full points are the median values from the distribution of the best-fit values. The data points of NGC 2434 and NGC 4486 are above the plot limits.} 
\label{fig8}
\end{figure}
\subsection{Matching models and observations}
\label{nomres}
We now compare the galaxy \ML\ gradient data, \dMLo, to the expectations from $\Lambda$CDM,
and illustrate this in Fig.~\ref{fig2},
where \dMLm\ is plotted versus $M_*$ for various values of $\epsilon_{\rm SF}$
and using the best-fit $R_{\rm e}$--$M_*$ relation ($\alpha=0.6, R_M=3.1$) together with the 68\% envelopes.
We can see immediately that the data do not seem to follow the simple increase of
\dML\ with $M_*$ that is expected for a ``universal'' $\epsilon_{\rm SF}$,
and the empirical increase appears to be sharper.
By inspection of the figure, one interpretation is that the low-mass galaxies 
have very high values for $\epsilon_{\rm SF}$ (in some cases even above the highest allowed limit) while the high-mass galaxies 
have a broad scatter in $\epsilon_{\rm SF}$, mostly within the allowed efficiency range. Note that projection effects in the dynamical
modelling of $\Upsilon$ can produce systematic effects on the inferred \dMLo: face-on disky galaxies could appear to have very low $\Upsilon_{\rm
out}$ if interpreted as spherical systems.  This would produce some data points with spuriously low \dMLo\ (and high $\epsilon_{\rm SF}$) values,
but these should comprise a small fraction of a large, unbiased sample.

To quantify the $\epsilon_{\rm SF}$--$M_*$ trend
requires allowing for the individual galaxies' departures
from the mean $R_{\rm e}$-$M_*$ relation.  
To this end, we fit the model simultaneously to $R_{\rm e}$ 
and to \dMLo\ with free parameters $(\alpha,
R_M, \epsilon_{\rm SF})$, minimising the following $\chi^2$ statistic:
\begin{equation} 
\chi^2=\sum_{i=1}^2\left(\frac{q_i-F_i}{\delta q_i}\right)^2 , 
\end{equation} 
where $q_1=R^{\rm obs}_{\rm e}$, $F_1=R_{\rm M}(M_*)^\alpha$,
$q_2$=\dMLo, $F_2=$\dMLm$(\alpha,R_{\rm M},\epsilon_{\rm SF})$ and
$\delta q_i$ are the uncertainties on $q_i$. 
\begin{figure} 
\vspace{-0.3cm}
\hspace{-0.5cm}
\epsfig{file=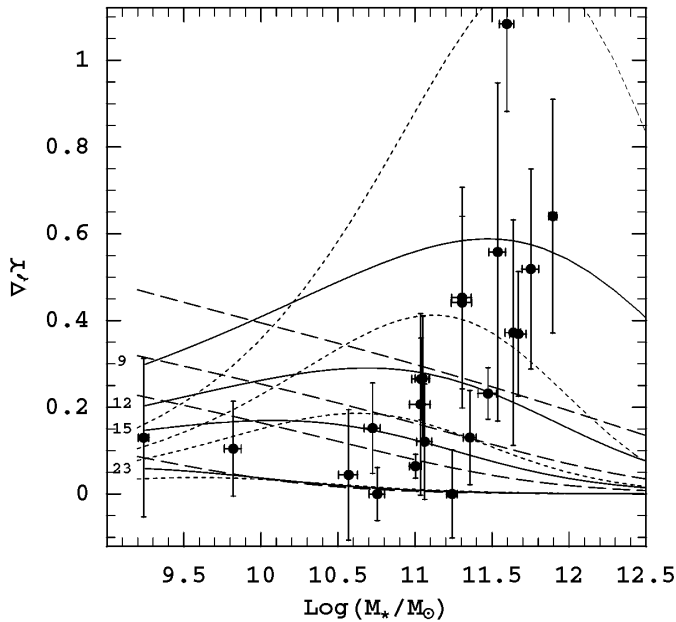,width=9cm,height=9cm,angle=-0}
\vspace{-0.65cm}
\caption{
\ML\ gradients versus total luminous mass, with curves shown of
constant concentration parameter $c_{\rm d}$, according to the
$\Lambda$CDM predictions, with $\epsilon_{\rm SF}$ free to vary. Line
styles are as in Fig. \ref{fig2}.
} 
\label{fig3}
\end{figure}  
\begin{figure} 
\hspace{-0.6cm}
\epsfig{file=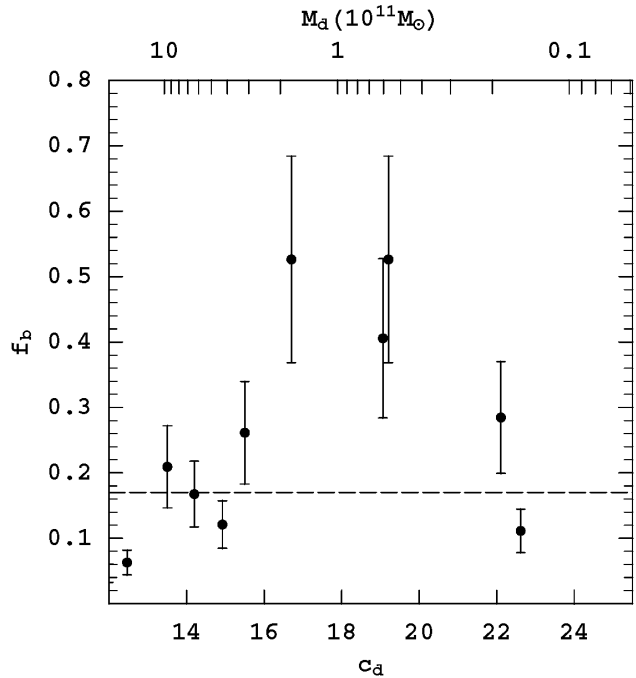,width=9.cm,height=9cm,angle=0}
\caption{
Baryon fraction for the low-mass systems ($M_*<M_0$) as a function of the halo concentration $c_{\rm d}$: here the high $c_{\rm d}\gsim 13$ range is shown. Halo masses are inferred by their best fit $c_{\rm d}$ according to the $c_{\rm d}-M_{\rm d}$ and shown on the upper axis. Dashed line marks the cosmological baryon fraction, $f_{\rm b}=0.17$.} 
\label{fig3a}
\end{figure} 
The results are shown in
Fig.~\ref{fig8} where we plot the best-fit $\epsilon_{\rm SF}$ (and
equivalently $f_{\rm d}$) against the stellar masses. From the distribution of the fitted values we have inferred the following median values: 
$\epsilon_{\rm SF}\sim 2.0_{-0.8}^{+4.0}$ for $\log M_*=11.0_{-0.4}^{+0.2}$ and $\epsilon_{\rm SF}\sim 0.23_{-0.13}^{+0.17}$ for $\log M_*=11.4_{-0.2}^{+0.3}$ (with a marginal overlap of the mass ranges around $\log M_*\sim11.2$). There are also galaxies with $\log M_*=11.7_{-0.1}^{+0.1}$ for which $\epsilon_{\rm SF}\sim 0.04_{-0.04}^{+0.03}$, only marginally consistent with the plausible lowest efficiency limit. These values suggest a smooth relation between $M_*$
and $\epsilon_{\rm SF}$ (see Fig. \ref{fig8}) which is at this stage only tentative because of the large uncertainties in the data.\\
An alternative interpretation is the presence of a mass
scale [$\log (M_0/M_{\odot})\sim 11.2$, $M_0 \sim 1.6\times10^{11}
M_{\odot}$] marking a clear dichotomy: galaxies with $M_* \gsim M_0$ are
most consistent with $\epsilon_{\rm SF}\sim 0.25$ $(f_{\rm d}\sim 19.5)$,
while galaxies with $M_* \lsim M_0$ are most consistent with $\epsilon_{\rm
SF} \sim $2 ($f_{\rm d}\sim 3$). In the latter sample, half of these are well
determined to have $\epsilon_{\rm SF} > 0.9$ $(f_{\rm d} < 5.4)$, while those galaxies with lower $\epsilon_{\rm SF}$ (larger \dMLo\ in Fig. \ref{fig2}) are {\em boxy} systems, in agreement with the tentative dichotomy discussed at the end of Section \ref{corr}. This suggests that there are separate populations of galaxies with very different $\epsilon_{\rm SF}$, rather than a trend of $\epsilon_{\rm SF}$ with $M_*$ for a single galaxy population---but we cannot distinguish between these two scenarios given the large uncertainties in  the current data. 

Note that other studies have suggested a maximum
galaxy formation efficiency (minimum dark-to-luminous mass fraction,
$f_{\rm d}$) at $M_*$ close to our $M_0$ \citep{ben00,mahu02,rom02}. Assuming
an average $\Upsilon_*=6.5$, as in our sample, we can convert this mass scale into a luminosity scale of $M_B=-20.5$, 
fairly similar to the luminosity scale found in other luminous properties of early-type galaxies \citep{grahal03,grahgu03}.\\ 
The $\epsilon_{\rm SF} \sim 0.25$ value for the high-mass galaxies is
logarithmically midway between the uppermost and lowermost limits adopted.
However, $\epsilon_{\rm SF}\sim2$ is unphysical for the low-mass galaxies,
given the assumptions of our $\Lambda$CDM models.  To understand the
origin of these $\epsilon_{\rm SF}$ values in the $\Lambda$CDM models, in
Fig. \ref{fig3} we plot \dMLm\ again as a function of $M_{\rm *}$ with
curves of constant $c_{\rm d}$ now indicated.  At each point on the curve,
$\epsilon_{\rm SF}$ has been adjusted to fit the required $\Lambda$CDM
$c_{\rm d}$-$M_{\rm d}$ relation. For a low stellar mass galaxy
($M_*<M_0$) to have a shallow \ML\ gradient, its $\Lambda$CDM concentration must be
high ($c_{\rm d}\gsim 15$), which implies a small halo mass. This mass corresponds to $\epsilon_{\rm SF}>1$ and thus to a
violation of baryon conservation.  To show this, we plot in
Fig.~\ref{fig3a} the baryon fraction $f_{\rm b}$ as a function of $c_{\rm d}$, where the $f_{\rm d}$ results (Fig. \ref{fig8}) have been re-interpreted with the
assumption $\epsilon_{\rm SF}=1.0$, giving $f_{\rm b}=1/(1+1.0f_{\rm d})$.
We find that many of the galaxies with $c_{\rm d}\gsim 15$ have
$f_{\rm b} \gg 0.17$, a circumstance which get worse if $\epsilon_{\rm SF}<1$.

\subsection{Discussion}
\label{poss}
\begin{figure} 
\hspace{-0.75cm}
\epsfig{file=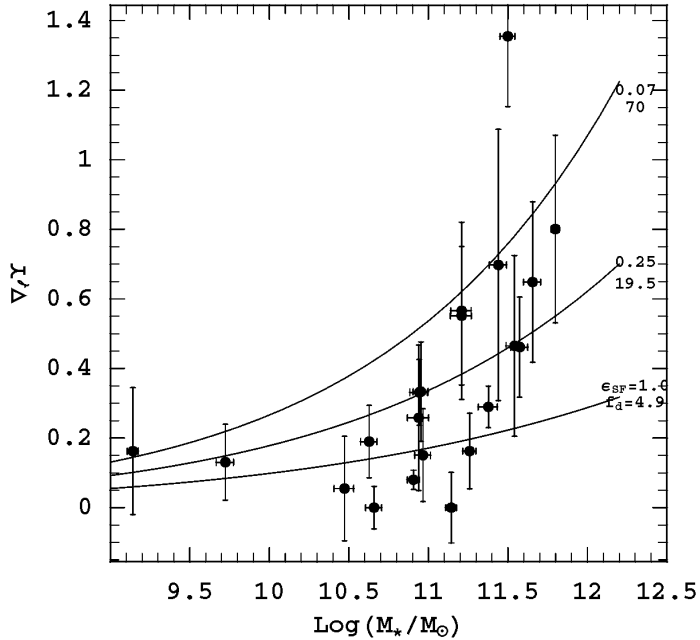,width=9.3cm,height=9.3cm,angle=-0}
 \vspace{-0.6cm}
\caption{
\ML\ gradients versus total luminous mass accounting for 30\% dark mass in the inner regions. Line styles have the same meaning as in Fig. \ref{fig2}. The position of the data points relative to the model curves is almost unchanged with respect to Fig. \ref{fig2}. 
\label{fig2b}} 
\end{figure}

In the previous section we have seen that the basic assumption of $\Lambda$CDM (namely the $c_{\rm d}-M_{\rm d}$ relation) and baryon conservation with $f_{\rm b}=0.17$ lead to typical formation efficiencies $\epsilon_{\rm SF}\sim 0.25$ for galaxies with $M_*>M_0$, while in the low-mass regime ($M_*<M_0$) they produce unphysical $\epsilon_{\rm SF}$. We have also concluded that if we want to decrease the inferred $\epsilon_{\rm SF}$ below 1.0, and still have $f_{\rm b}=0.17$, we must release the $c_{\rm d}-M_{\rm d}$ relation otherwise we produce a violation of the baryon conservation.\\
Before moving on to other possible explanations for this problem, we first examine whether this problem
can be explained by the systematic uncertainties in $\Upsilon_*$.
As discussed in Section~\ref{frac}, in the $\Lambda$CDM framework, we expect $\sim$10--20\% of $\Upsilon_{\rm in}$
to be a dark matter contribution.
Reducing $\Upsilon_*$ by as much as 30\% will systematically shift all the empirical estimates
of \dMLo\ higher by 30\% (see Eq.~\ref{grad0}).
The $R_{\rm e}$--$M_*$ relation will also be affected, such that $R_{\rm e}$ for a given $M_*$ is
higher, and thus $r_{\rm in}/r_s$ and $r_{\rm out}/r_s$ in the
model calculations will be higher;
this leads to a higher value of \dMLm\ predicted by the models which roughly offsets the
higher values in the data.
The final effect is for the data values of $M_*$ to shift, and so the net change is for
the low-mass galaxy data points in Fig.~\ref{fig2} to shift roughly parallel to the
contours of constant $\epsilon_{\rm SF}$---not affecting the conclusions on $\epsilon_{\rm SF}$ significantly
(see Fig.~\ref{fig2b}).
To decrease the inferred $\epsilon_{\rm SF}$ for the intermediate-mass galaxies to values below
1.0 would require a dark mass fraction inside $R_{\rm e}$
of $\gsim 50$\%.
Such high central dark matter fractions
are not supported
by other empirical studies of the central mass content of early-type galaxies
\citep{gerhard01,rusal03}.
They
would also imply that the average stellar mass-to-light ratios,
$\Upsilon_{B,*}=6.5\pm 2.0$, must be decreased by
a factor of two: this is not plausible from the standpoint
of stellar populations modelling where $\Upsilon_{B,*} \lsim 4$ only for fairly
young stellar populations, with age $\lsim 5$ Gyr, and metallicity $-0.5<[{\rm Fe/H}]<0.5$ \citep{W94,Mar98}.\\
One way to avoid the problem of low-mass haloes is
to relax the $c_{\rm d}$--$M_{\rm d}$ relation,
using a maximum plausible value of $\epsilon_{\rm SF}$ as a prior.
\begin{figure} 
\hspace{-0.6cm}
\epsfig{file=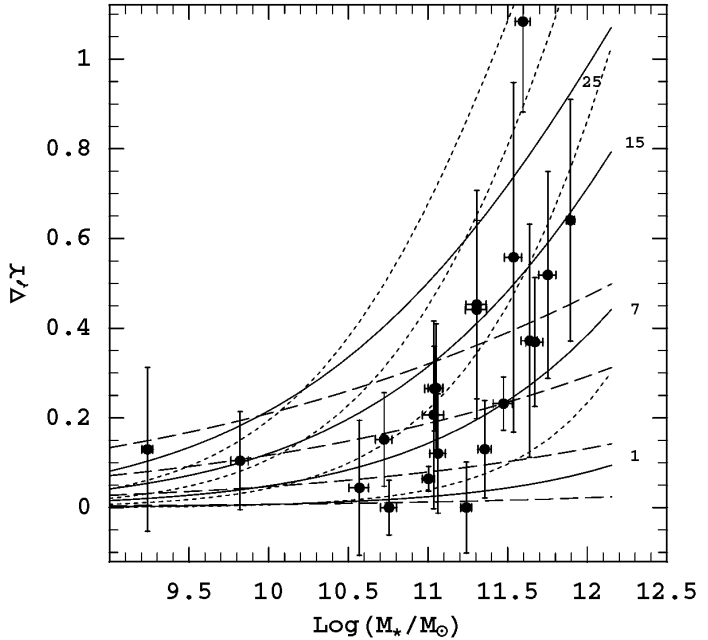,width=9.3cm,height=9.3cm,angle=0}
\vspace{-0.6cm}
\caption{\ML\ gradients versus total luminous mass and concentration parameter 
after $c_{\rm d}$--$M_{\rm d}$ relation relaxed for $\epsilon_{\rm SF}=0.25$ . 
Labelled solid lines are the predictions from the $\Lambda$CDM for different $c_{\mathrm{d}}$: line styles as in Fig. \ref{fig2}.
\label{fig3b}} 
\end{figure}   
Once $\epsilon_{\rm SF}$ and $M_{\rm *}$ are fixed, the total dark-halo mass, $M_{\rm d}$, is set and the only way to alter $M_{\rm d}(r)$ (Eq. \ref{MNFW}) is by altering $c_{\rm d}$.
In this way \dMLm\ can be written as a function of $M_{\rm *}$ and $c_{\rm d}$: in Fig.~\ref{fig3b} 
this is shown for $\epsilon_{\rm SF}=0.25$ ($f_{\rm d}=19.5$).
We can derive the best-fit $c_{\rm d}$ values for each galaxy using the same procedure adopted for best-fitting $\epsilon_{\rm SF}$ (as in previous section).\\
Results are in Fig. \ref{fig4} ({\it left}):
we see that the most massive galaxies ($M_{\rm d}\sim 8\times10^{12} M_{\odot}$)
have concentrations consistent with $\Lambda$CDM expectations
($c_{\rm d} \sim$~8--14),
while many of the less massive galaxies ($M_{\rm d}\sim 2\times10^{12} M_{\odot}$)
have very low concentrations ($c_{\rm d} \sim$~1--6).\\
The shallow \ML\ gradients in the low-mass galaxies are thus a simple consequence
of a small fraction of the dark matter residing in the galaxy centre.
This explanation was advanced in \cite{ral03} but contradicts $\Lambda$CDM expectations
that $c_{\rm d}$ is decreasing with $M_{\rm d}$ rather than increasing.
As suggested by Section \ref{nomres}, $\epsilon_{\rm SF}$ may
not be constant but rather decreasing with galaxy mass, which may alleviate the problem.  We thus assume $\epsilon_{\rm
SF}$=0.25 for $M_* > M_0$ and $\epsilon_{\rm SF}=1.0$ for $M_* < M_0$
(a maximum plausible value), and plot the results for $c_{\rm d}$ in Fig. \ref{fig4} ({\it right}).
The $c_{\rm d}$--$M_{\rm d}$ relation is now in a better agreement with the $\Lambda$CDM expectation,
but the low-mass galaxies still have haloes with concentrations which are too low on average.

The above alternatives (baryon non-conservation; low concentrations) for
producing low \ML\ gradients are necessitated by the strict NFW
functional form $M_{\rm d}(r)$ for the haloes (Eq.~\ref{MNFW}).
Thus there is little flexibility to change 
$M_{\rm d,out}$ vs $M_{\rm d,in}$.
But another way to
reduce \dML\ in the models (Eq. \ref{grad})
is to change the ratio of $M_{\rm d,out}$ versus $M_{\rm d,in}$
by changing the form of the mass profile $M_{\rm d}(r)$.
For example, in the classic adiabatic contraction model of galaxy formation, the baryons' collapse
drags dark matter inward; this will increase $M_{\rm d,in}$ above the nominal NFW halo central mass
and decrease the inferred \dMLo.  
Again, the implications of this alternative would be that many other
studies of central DM density are incorrect,
and that the population synthesis models have over-estimated $\Upsilon_*$ by $\sim$50\%.
As mentioned in Section~\ref{cdm}, it is not clear that this adiabatic contraction scenario is really to be expected,
and we leave examination of the details to a future study---but comparing the alternatives, it seems
this may be the most likely one for ``saving'' $\Lambda$CDM.

Of course, the $\Lambda$CDM theory is not as readily falsifiable as may be supposed, as various modifications
to its simplest outline had been proposed in order to solve various problems.
One may assume a different initial power spectrum and in principle partially solve the problems seen here;
it has also been suggested that $\sigma_8 \sim 0.75$ is a reasonable solution to the low-concentration halo problem
\citep{mcgaugh03,vdB}.
\begin{figure*}
\hspace{-1.5cm}
\epsfig{file=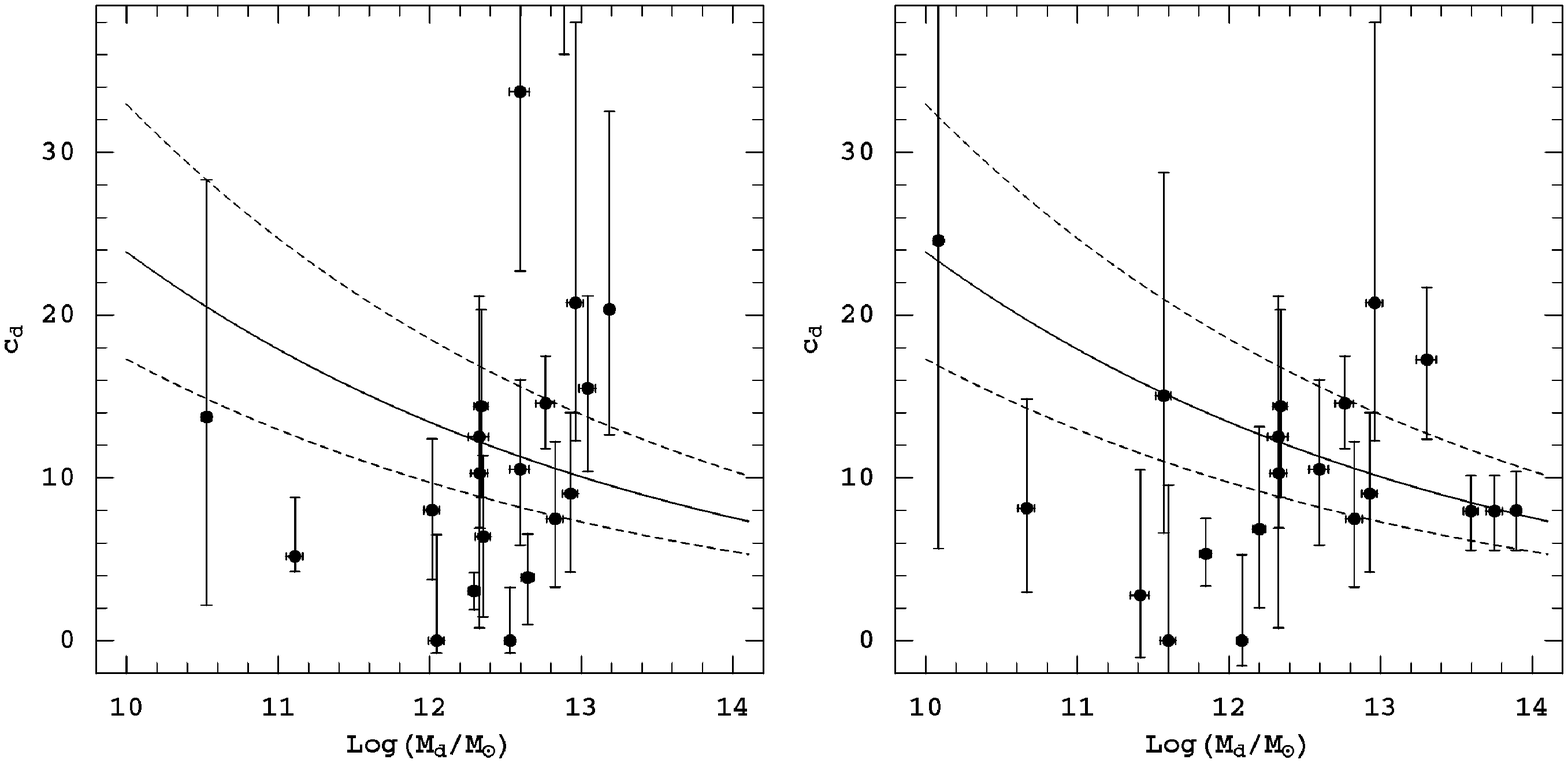,width=18cm,height=9cm,angle=0}
\caption{
Best fit $c_d$ values after relaxing the $c_{\rm d}$--$M_{\rm d}$ relation for $\epsilon_{\rm SF}=0.25$ (left) and $\epsilon_{\rm SF}$ varying within the galaxy sample (right: see details in the text). Solid line is $c_{\mathrm{d}}-M_{\mathrm{d}}$ from B01 with 1-$\sigma$ scatter indicated by dashed lines.  
\label{fig4}} 
\end{figure*} 
\subsection{Comparison with other dynamical techniques}\label{otech}
For the sake of completeness, in this paragraph we briefly discuss how the results based on stellar kinematics plus PNe and GCs compare with results listed in the literature based on the techniques we have excluded in this paper: namely X-rays, gravitational lensing and HI rings. This is to verify whether (regardless any possible selection effects) different techniques give consistent results for the dependence of the \ML\ gradients on the stellar mass and luminosity, which we have explained as a trend of the star formation efficiency with the same global quantities.\\ 
~\\
{\em X-rays.} \citet{OsPon04b} reviewed literature results on \ML\ estimates based on X-ray emissions in early-type galaxies. They found that at 5--6 $R_{\rm e}$ different modeling techniques give a lower limit of \ML=20 $M_{\odot}/L_{\odot}$ on a large galaxy sample. Even if most of these estimates come from cluster and group dominant galaxies, they concluded that ellipticals can possess dark haloes of their own, regardless of their environment. We can infer a lower limit for the \dMLo\ for these systems using their results. We can assume a $\Upsilon_{\rm in}=7$ as the average value find in our sample and derive \dMLo(Xrays)$\ge0.4$. Since $M_B\lsim -20.5$ in the X-rays sample, this is consistent with $\epsilon_{\rm SF}\lsim0.25$ found for our galaxy sample of stellar masses $M_*>M_0$.\\ 
~\\
{\em Strong Gravitational lensing (SGL).} There are only a few studies where \ML\ of the innermost and outer regions are available for reliable \dMLo\ estimates. 
As discussed above, there is the further complication that SGL analyses refer to high $z$ ($z\geq 0.1$) 
and they cannot be directly compared to model predictions obtained in the local universe ($z=0$) 
since both dark and luminous matter scaling relations change with redshift.
A detailed model prediction at high--$z$ is beyond the purpose of this paper. Here we want to give only a qualitative interpretation of SGL results.

\citet{TrKoo04} reviewed 5 bright ($M_B\lsim -20.5$) E/S0 systems
with reliable \ML\ estimates inside the Einstein radius (ranging from 0.8 to 5 $R_{\rm e}$). 
Using their results we compute \dMLo$\geq 0.4$ and an average \dMLo(SGL)~$\sim0.8$.
These values are consistent with but somewhat higher than our dynamical results
for bright galaxies, and
comparison to $z=0$ models 
implies a very 
low star formation efficiency ($\epsilon_{\rm SF}\lsim0.1$). 

\cite{rusal03} used statistical constraints on the mass profiles of
22 lens systems over a broad range of luminosities to find a typical
homologous mass profile.
Their results assuming a $\Lambda$CDM dark matter profile imply
$\dML \sim 0.1$--$0.2$, which is consistent with our results for faint
galaxies but apparently not consistent with the \cite{TrKoo04} results
and with our bright galaxy results.
A more careful examination of these results, including treatment of
the $z$-dependence of the models,
awaits a future paper.\\
~\\
{\em HI rings.} For these galaxies, the gradient prediction should account the presence of the gas since its mass contribution it is not negligible for these systems. For the sake of simplicity, as a first approximation we use our predictions obtained without gas as a reference. There are a few systems in the literature with reliable innermost and outer \ML\ values suitable for \dMLo\ estimates.\\
These galaxies include NGC 5266 \citep{morg97}, IC 5063 \citep{morg98},
IC 2006 \citep{franx94}, NGC~1052, NGC 2974 and NGC 4278 \citep{bertola}.
These galaxies span a broad range of luminosity, and we derive \dMLo\ values
ranging from 0.4 to 1.0,
corresponding to $\epsilon_{\rm SF} \lsim 0.2$.
These results are consistent with what we have found for the brighter galaxies,
but the weak halo behaviour for fainter galaxies is not detected.
However, as discussed earlier, we consider it dangerous to take the HI results as typical
of early-type galaxies because of selection issues.

\subsection{Dwarf galaxies}\label{lowmass}
In Section \ref{mod} we have made \dML\ predictions for a wide range of galaxy masses, including dwarf systems  ($\log M_* \lsim 10$). This mass regime is populated both by dwarf elliptical galaxies (dEs) and low surface brightness late-type systems (LSBs). These two classes are different in many respects (morphology, gas content, stellar population etc.), but they also share some similarities: they are small, low-luminosity galaxies with diffuse, exponential declining surface brightness profiles\footnote{In this respect our basic model assumption of an $R^{1/4}$ surface brightness profile is poorly representative for these systems.}. Furthermore they have the same size-mass relation \citep{shen03} and hence they are expected to have similar dark-to-luminous mass distributions. Specifically, as shown in Section \ref{pred}, we must expect observed gradients to increase for decreasing stellar masses (and luminosities) and for decreasing star formation efficiency (at any given mass).
Unfortunately, these galaxies remain among the most poorly studied systems, mostly because of the observational difficulties related to their faint luminosities and detailed dynamical analysis are scarce and generally focussed on the central galaxy regions \citep{beni90,mateo98}.\\
A complete view of these systems, although beyond the purpose of this paper, is complicated by data which are generally not suitable for the kind of analysis based on \ML\ gradients. 
Here we want to briefly compare our models with a few literature data sets which can be adapted to this analysis. We restrict our analysis to dwarf early-type galaxies which are gas poor and can be directly compared to the predictions of Section \ref{mod}, where we discarded any gas component.

{\em Fornax} \citep{walch03} [$M_B$=-11.7]: they estimate \ML=23 at 90$'$ from the galaxy centre ($R_{\rm e}=9.9'$) and a central \ML=4.8. We have obtained \dMLo=0.52 for this galaxy which corresponds to $\epsilon_{\rm SF}$=0.6 using the dwarf shallower $R_{\rm e}-M_*$ relation in our models. We found consistent $\epsilon_{\rm SF}$ values using \ML\ estimates from \citet{loka01}.

{\em FS373} \citep{Derij04} [$M_B$=-16.9]: they obtained \ML=7.8 at 1.5 $R_{\rm e}$(=7.9$''/$1.57 kpc) and computed  a stellar \ML=2--4 from age and metallicity of the stellar population. This allowed us to estimate  \dMLo$\sim$1--2.5 and $\epsilon_{\rm SF}$=0.1--0.6.

{\em FS76} \citep{Derij04} [$M_B$=-16.1]: they obtained \ML=7.4 at 1.5 $R_{\rm e}$(=4.4$''/0.77$ kpc) and considered a stellar \ML=2--4 as for FS373. We found pretty similar \dMLo$\sim$1--2.5 but $\epsilon_{\rm SF}$=0.2--0.45 for this galaxy.\\

This very preliminary test on dwarf ellipticals suggests that \ML\ gradients are actually increasing with decreasing masses as expected, and that this trend is also consistent with $\epsilon_{\rm SF}$ 
tending to decrease at this mass scale from a maximum near $L^*$.\\ 
Low star formation efficiencies in this mass range have been also inferred for gas-rich LSB galaxies \citep{joca90,cote+91,mart+94}. They found $f_{\rm d}\sim M_{\rm tot}/M_*=$25--27\footnote{$M_{\rm lum}$ in their paper refers to the total baryonic matter $M_{\rm (stars + gas).}$} in the galaxies NGC 3109, NGC 5855 and IC 2574 respectively, corresponding to observed $\epsilon_{\rm SF}\sim0.2$ which are almost in agreement with efficiencies of dwarf ellipticals derived above. \citet{lake+90} claim at least half of the baryonic mass of DDO 170 in the form of gas, which implies $\epsilon_{\rm SF}\lsim0.5$ assuming baryon conservation.
All these $\epsilon_{\rm SF}$ estimates are consistent with 
what we obtain
by comparing the \dMLo\ from their \ML\ estimates with our models (in spite of our model simplification of not including gas).\\ 
We can then conclude that, based on \ML\ gradients, there is evidence of low star formation efficiencies in the very low mass regime. This seems in agreement with other studies \citep{ben00,mahu02} claiming a continuous decreasing star forming efficiency with decreasing masses at these mass scales.

\section{Summary and Conclusions} 
We have introduced a method for measuring \ML\ gradients in early-type (elliptical and lenticular) 
galaxies, quantified by \dML.
If the same technique is used for all \ML\ measurements in a galaxy,
then \dML\ is independent of bandpass definitions and distance uncertainties.
We have built typical galaxy models in the $\Lambda$CDM framework using NFW profiles for
the dark matter distribution.
Assuming baryon conservation across the virial radius, these galaxies are parametrised
by their total stellar mass $M_*$ and their net star formation efficiency $\epsilon_{\rm SF}$;
we have then made predictions for \dML\ (between 0.5~$R_{\rm e}$ and 4~$R_{\rm e}$) 
as a function of $(M_*, \epsilon_{\rm SF})$.
As a consequence of the relative spatial scaling relations of the luminous and dark matter,
we have found that brighter galaxies should typically show higher values of \dML\ than 
fainter ones, {\it appearing} more dark-matter dominated even though the overall dark-to-luminous
mass fraction $f_{\rm d}$ is a constant.
These predictions are valid for a mean set of galaxies only, and should not be used to draw
direct conclusions about individual galaxies, whose properties are subject to considerable scatter.

We have next assembled a set of \ML\ measurements (using stellar, PN and GC kinematics) in 21 early-type galaxies as a pilot study
for comparing to our model predictions.
We have found that \dML\ in these galaxies increases with $M_*$ (and with other correlated parameters), 
which is evidence for systematic changes in either the fraction of dark matter or in its
radial distribution.
This empirical trend must be explained by any successful model of
galaxy properties and formation 
(including those advocating alternative theories of gravity). Environmental density does not appear to play a key role in producing these trends.

Comparing the empirical \dML\ values to the
$\Lambda$CDM predictions, we find a stellar mass scale $M_0 \sim 1.6 \times
10^{11} M_{\odot}$, and a corresponding luminosity scale of $M_B \sim
-20.5$, marking a galaxy dichotomy.
The brighter (more massive) galaxies
$(M_B \lsim -20.5)$ show a broad scatter in $\epsilon_{\rm SF}$, but are generally consistent
with $\epsilon_{\rm SF}\sim$~0.25.
The fainter (low-mass) galaxies are generally only consistent with $\epsilon_{\rm SF}\sim$~2--3, which is
not physically possible in the context of our model assumptions.\\
The brighter galaxies thus appear consistent with the $\Lambda$CDM picture
without need for further complications from baryonic effects.
For the fainter galaxies, we have examined some possibilities for explaining their 
apparent extremely high $\epsilon_{\rm SF}$ values
(i.e., apparent extremely low dark matter fractions).
One explanation is that the baryon fraction has not been conserved in these galaxies:
either they have acquired large quantities of extra baryons from inflows across the virial radius
(e.g., gas from nearby starbursting galaxies, which has since formed new stars); 
and/or they have lost portions of their outer dark haloes 
(after star formation) through dynamical interactions with other galaxies and groups.
We note that the latter process is implicitly included in $\Lambda$CDM simulations of
dark halo formation, but the net effect on $f_{\rm d}$ has not been explicitly studied.

Another explanation is that the predicted relation between halo mass $M_{\rm d}$
and concentration $c_{\rm d}$ in $\Lambda$CDM is wrong, and 
$c_{\rm d}$ tends to {\it increase} rather than {\it decrease} with $M_{\rm d}$,
with $c_{\rm d}\sim$~1--6 for $L^*$ galaxies.
Similar conclusions have already been reached in studies of low-surface brightness galaxies
\citep{alam,mcgaugh03}, and may indicate that the dark matter is not CDM,
or that the baryonic processes involved in galaxy formation have somehow caused the
dark haloes to become less centrally-concentrated.
However, the current best guess for baryonic effects is that the cooling and collapse
of the gas causes the dark matter halo to also contract adiabatically
\citep{blum86,gne04}, which {\it increases} the central density of the halo.
This is still a viable scenario for the fainter early-type galaxies if the stellar
\ML\ values are systematically lower by 50\% than what we have assumed,
and the central dark mass fraction is $\gsim$40\%.  
This possibility merits more investigation, although we note that several studies have concluded
that the dark matter cannot be so dominant.
Both scenarios (low-concentration haloes; low stellar \ML)
are possible ways for resolving an apparent 
empirical inconsistency:
weak lensing and satellite dynamics studies find
evidence for large amounts of dark matter inside the virial radius in $L^*$ 
early-types \citep{guse02,vdBal04},
while this paper suggests little dark matter 
inside $\sim 4 R_{\rm e}$ in similar galaxies.

Whatever the case, these data imply a dichotomy in the dark halo profiles between
fainter and brighter early-type galaxies (see Section \ref{corr}), which parallels the dichotomy already
visible in the properties of their luminous bodies \citep{nieto,cap92,fab97,grahal03,grahgu03}, and X-ray properties \citep{pel99},
and suggests markedly different formation histories for the two galaxy populations where baryon cooling efficiencies could have played a pivotal role.
The comparison of data and models also supports other studies indicating that
$\epsilon_{\rm SF}$ is at a maximum for galaxies around $L^*$
\citep{ben00,guse02,mahu02,vdB}. Indeed, we have briefly discussed the case of dwarf/LSB galaxies and shown that there is the evidence of decreasing $\epsilon_{\rm SF}$ from their \dMLo, supporting this scenario.
We do not wish to over-interpret the implications for $\Lambda$CDM at this point, since
the number of fainter galaxies in the sample is not yet statistically large,
and most of them are not yet modelled in enough detail to exclude the potentially
important effects of radial anisotropy at large radius.
But this paper highlights the importance of detailed dynamical studies of a larger
sample of galaxies, where the distribution of the dark matter as well as the stars may be
used as clues to the formation histories of galaxies.
\section*{Acknowledgments}
We are very grateful to the referee, Dr. A. Bosma, for his comments which allowed us to improve the paper results. We thank James Bullock for providing his toy model code. NRN thanks M. Pannella for stimulating discussion on galaxy formation. NRN is receiving a grant from the EU within the 5th Framework Program (FP5) -- Marie Curie Individual Fellowship. NRN, MC and MA acknowledge financial support by INAF-Project of National Interest.

\end{document}